# Analytic guiding center formulas for bounce-transit motion in a concentric circular, finite inverse aspect ratio tokamak geometry

Cole D. Stephens,[1,2,a)] Xavier Garbet,[3] and Frank Jenko[2]

## AFFILIATIONS

[1]University of California, Los Angeles, 475 Portola Plaza, Los Angles, California 90095, USA
[2]Max Planck Institute for Plasma Physics, Boltzmannstr. 2, 85748 Garching, Germany
[3]CEA, IRFM, F-13108 Saint Paul-lez-Durance, France

[a)]Author to whom correspondence should be addressed: cdstephens@physics.ucla.edu

## ABSTRACT

Bounce-transit motion in concentric circular magnetic geometry is typically analyzed in the limit that the inverse aspect ratio, $\varepsilon$, is small. We prove that this approximation is not necessary to study a concentric circular geometry by deriving new analytical formulas while retaining a non-zero $\varepsilon$. We use these formulas to demonstrate that the approximation is robust for $\varepsilon \lesssim 0.3$.

## I. INTRODUCTION

Inhomogeneity in a magnetic field gives rise to charged particle orbits such that the particle either is trapped in the magnetic well or passes through the magnetic field.[1–3] The properties of these orbits have long been studied in tokamak plasmas and are instrumental in describing a wide range of kinetic and neoclassical phenomena.[4–10] For example, analysis of trapped electron modes requires knowledge of the lowest-order bounce-transit motion, and the toroidal drift frequency characterizes the stability of these modes.[11–13] In kinetic magnetohydrodynamics, the bounce-transit and drift frequencies underlie the resonances with energetic particles that drive instabilities.[14–16] Moreover, neoclassical calculations of the zonal flow residual involve a series of bounce-transit time averaging, a process that is intimately related to the calculation of the toroidal drift frequency.[17–19] Thus, analytical formulas are indispensable in understanding these problems through analytical and computational means; this need has led to a great amount of work done solely on the bounce-transit motion itself in order to obtain such formulas.[2,20] Unfortunately, the complex magnetic geometry in a tokamak presents a difficult obstacle in deriving analytical formulas. The main goal of this work is to remove one of these obstacles to allow for more accurate calculations.

Previous analyses of guiding center bounce-transit motion have utilized concentric circular magnetic geometry to simplify the problem. In addition, the inverse aspect ratio is typically assumed to be small, resulting in an even more approximate magnetic geometry.[12,20,21] This work takes direct inspiration from Ref. 20, which performed analogous calculations in the limit of small inverse aspect ratio. While this small inverse aspect ratio approximation does make the formulas easier to obtain and more compact, they are not necessary to carry out an analytical derivation. Instead, a finite, non-zero inverse aspect ratio can be retained throughout the entirety of the derivation. The calculations presented in this work are possible because the fundamental integrals used in this paper are of the form

$$\int R(u, \sqrt{P(u)})du, \tag{1}$$

where $P$ is a polynomial of degree 3 or 4 with no repeated roots, $R$ is a rational function of $u$ and $\sqrt{P(u)}$, and $R$ contains at least one odd power of $\sqrt{P}$.

It is well known that such an integral can always be represented by the three Legendre forms of elliptic integrals.[22–25] While previously derived formulas only used elliptic integrals of the first and second kind, these new formulas require the elliptic integral of the third kind. Closed form solutions for the field line-following equations of motion are found using modified functions similar to the classical Jacobi elliptic functions that are used, for instance, to describe particle trapping in waves.[26] We also show that the small inverse aspect ratio approximation can be directly applied to the finite inverse aspect ratio analysis.

The resulting comparison demonstrates that the approximation is valid for a moderately small inverse aspect ratio. In the cases where the approximation is not good enough, these analytical formulas can be used for increased accuracy. Spherical tokamaks such as the National Spherical Torus (NSTX) and the Mega Ampère Spherical Tokamak (MAST) are characterized by low aspect ratios; the inverse aspect ratios for these two tokamaks were both $\sim 0.77$.[27,28] Thus, finite inverse aspect ratio effects can be important, even if the analysis is restricted to a concentric, circular geometry.

The paper is organized as follows. Section II clarifies the analytical concentric circular magnetic geometry and the preliminaries for bounce-transit motion along the field line. Sections III and IV derive formulas for the bounce-transit frequency, the longitudinal adiabatic invariant, and the closed form solution to the bounce-transit motion. We next take into account deviations from the field line motion due to magnetic drifts and conservation of canonical toroidal momentum while still retaining a finite inverse aspect ratio. Section V analyzes the radial excursion from the magnetic field line that results in banana orbits. Section VI then calculates the slow toroidal drift frequency by analyzing the equation of motion in the toroidal direction. Finally, we summarize our work in Sec. VII. We also include a brief overview of our definitions and notation for elliptic integrals in Appendix A, while a tabulation of specific elliptic integrals used to calculate the toroidal drift frequency can be found in Appendix B.

## II. CIRCULAR MAGNETIC GEOMETRY

For a concentric circular tokamak geometry,[29] the magnetic field is defined to be

$$\mathbf{B} = \frac{B_0}{1+\varepsilon\cos(\theta)}\left(\hat{\phi} + \frac{\varepsilon}{\bar{q}(r)}\hat{\theta}\right). \tag{2}$$

The triad $(r, \theta, \phi)$ (minor radial distance, poloidal angle, azimuthal angle) defines a right-handed coordinate system with $\hat{\mathbf{r}} \times \hat{\theta} = \hat{\phi}$. The inverse aspect ratio $\varepsilon = r/R_0$ and approximate safety factor $\bar{q}$ are functions of $r$ only, where $R_0$ is the major radius. We do not take into account the Shafranov shift; thus, the flux surfaces considered are concentric circles.[30] We note that the magnetic field strength is typically approximated as

$$B = |\mathbf{B}| \approx B_0(1-\varepsilon\cos(\theta)). \tag{3}$$

To retain a finite inverse aspect ratio, we do not use this approximation in this work.

The true safety factor $q(r)$ is given by

$$q(r) = \frac{1}{2\pi}\int_0^{2\pi} d\theta\, \frac{\mathbf{B}\cdot\nabla\phi}{\mathbf{B}\cdot\nabla\theta} = \frac{\bar{q}(r)}{\sqrt{1-\varepsilon^2}}. \tag{4}$$

We note that the magnetic field can be written in the Clebsch form[31]

$$\mathbf{B} = \nabla\alpha \times \nabla\psi, \tag{5}$$

where

$$\alpha = \phi - q(r)\chi(r,\theta), \tag{6}$$

$$\frac{d\psi}{dr} = \frac{rB_0}{\bar{q}(r)}. \tag{7}$$

We identify $\psi$ with minus the poloidal flux normalized to $2\pi$ and note that $\psi$ labels the flux surface while $\alpha$ labels the field line. The straight field line poloidal coordinate $\chi$ is defined as

$$\chi = 2\arctan\left[\sqrt{\frac{1-\varepsilon}{1+\varepsilon}}\tan\left(\frac{\theta}{2}\right)\right], \tag{8}$$

and is constructed to satisfy the property

$$\frac{\mathbf{B}\cdot\nabla\phi}{\mathbf{B}\cdot\nabla\chi} = q(r). \tag{9}$$

The unit vector $\hat{\mathbf{b}}$ denoting the direction of the magnetic field is

$$\hat{\mathbf{b}} = \frac{\mathbf{B}}{B} = \frac{1}{\sqrt{1+\varepsilon^2/\bar{q}^2}}\left(\hat{\phi} + \frac{\varepsilon}{\bar{q}}\hat{\theta}\right). \tag{10}$$

The guiding center approximation used in this work is justified by the adiabatic conservation of the magnetic moment, which requires the gyroradius to be small compared to other relevant length scales in the system.[32–36] We impose two conditions on the guiding center particle orbit to determine the bounce-transit motion to lowest-order in the gyroradius

$$\dot{\psi} = 0, \tag{11}$$

$$\dot{\alpha} = 0. \tag{12}$$

This restricts the particle to the magnetic field line. These conditions can be written in terms of our previous coordinates,

$$\dot{r} = 0, \tag{13}$$

$$\dot{\phi} = \frac{\bar{q}\dot{\theta}}{1+\varepsilon\cos(\theta)}. \tag{14}$$

We then compute the parallel velocity with this restriction in mind,

$$v_\parallel = \mathbf{v}\cdot\hat{\mathbf{b}} = \sqrt{1+\frac{\varepsilon^2}{\bar{q}^2}}\,\bar{q}R_0\dot{\theta}. \tag{15}$$

Accordingly, we can define the length along the field line to be

$$l = \sqrt{1+\frac{\varepsilon^2}{\bar{q}^2}}\,\bar{q}R_0\theta. \tag{16}$$

From conservation of energy, the magnitude of the parallel velocity is

$$|v_\parallel| = \sqrt{\frac{2}{m}(E-\mu B)} = \sqrt{\frac{2E}{m}}\sqrt{1-\frac{\mu B^0/E}{1+\varepsilon\cos(\theta)}}, \tag{17}$$

where $E$ is the kinetic energy (since we ignore the electrostatic potential), $\mu$ is the magnetic moment, and

$$B^0(r) = B_0\sqrt{1+\varepsilon^2/\bar{q}^2}. \tag{18}$$

We next parameterize the orbit via the trapping parameter $\kappa$,

$$\frac{\mu B^0}{E} = 1+\varepsilon-2\varepsilon\kappa^2. \tag{19}$$

Trapped particles reflect at a poloidal angle $\theta_b$ where their parallel velocity is 0. We find that for trapped particles,

$$0 \le \kappa < 1, \tag{20}$$

$$\theta_b = 2\,\arcsin(\kappa), \tag{21}$$

Passing particle orbits, however, are not reflected. Instead, they continue along the magnetic field line and make a complete transit through the entire torus. Thus, the trapping parameter for passing particles is bounded by

$$1 < \kappa \le \sqrt{\frac{1+\varepsilon}{2\varepsilon}}. \tag{22}$$

The upper bound for $\kappa$ corresponds to the situation where the perpendicular velocity is 0. A particle at the trapped-passing boundary ($\kappa = 1$) is analogous to a stationary pendulum at the top of its arc, so we expect the period to approach infinity as $\kappa \to 1$ from both sides. We note that in the small inverse aspect ratio limit, the upper bound for $\kappa$ approaches infinity.

The differential equation governing the motion of the particle is then

$$\sqrt{1+\frac{\varepsilon^2}{\bar{q}^2}}\bar{q}R_0|\dot{\theta}| = \sqrt{\frac{2E}{m}}\sqrt{2\varepsilon}\sqrt{\frac{\kappa^2-\sin^2(\theta/2)}{1+\varepsilon\cos(\theta)}}. \tag{23}$$

The characteristic frequency for the periodic motion is

$$\omega_0 = \frac{\sqrt{E/m}\sqrt{\varepsilon}}{\bar{q}R_0\sqrt{1+\varepsilon^2/\bar{q}^2}}, \tag{24}$$

while the characteristic length is

$$l_0 = \sqrt{1+\frac{\varepsilon^2}{\bar{q}^2}}\bar{q}R_0. \tag{25}$$

This gives us

$$|\dot{\theta}| = 2\omega_0\sqrt{\frac{\kappa^2-\sin^2(\theta/2)}{1+\varepsilon\cos(\theta)}} \tag{26}$$

We confirm that in the small $\varepsilon$ limit

$$\omega_0 \approx \frac{\sqrt{E/m}\sqrt{\varepsilon}}{\bar{q}R_0}, \tag{27}$$

$$l_0 \approx \bar{q}R_0. \tag{28}$$

Having defined the exact concentric circular geometry and identified the lowest-order equation of motion, we proceed to deriving analytical formulas for the bounce-transit frequency and the longitudinal adiabatic invariant.

## III. BOUNCE-TRANSIT FREQUENCY

Although the guiding center deviates from the magnetic field line, we can calculate the longitudinal invariant $J$ (also known as the bounce-transit action); $J$ is an adiabatic invariant, meaning that if the timescale of the deviation is longer than the bounce-transit period then it can be treated as a constant.[37] $J$ is defined to be

$$J = \frac{1}{2\pi}\oint m v_{\parallel}\,dl. \tag{29}$$

Meanwhile, the bounce-transit frequency is defined as

$$\omega = \left(\frac{\partial J}{\partial E}\right)^{-1}, \tag{30}$$

or equivalently

$$\omega = \frac{2\pi}{T} = 2\pi\left(\oint \frac{dl}{v_{\parallel}}\right)^{-1}. \tag{31}$$

Because the bounce and transit motions are qualitatively different, we separate the calculation for trapped and passing particles. This distinction will apply to all other calculations in this paper.

The elliptic integrals calculated here and elsewhere in the paper are functions of parameters called the modulus and the characteristic. These parameters are functions of $\kappa$ and $\varepsilon$ and take on different forms for trapped and passing particles. For trapped particles, the modulus and characteristic are

$$k_b = \kappa\sqrt{\frac{1-\varepsilon}{1+\varepsilon-2\varepsilon\kappa^2}}, \tag{32}$$

$$n_b = \frac{-2\varepsilon\kappa^2}{1+\varepsilon-2\varepsilon\kappa^2}. \tag{33}$$

For passing particles, the modulus and characteristic are

$$k_t = \frac{1}{\kappa}\sqrt{\frac{1+\varepsilon-2\varepsilon\kappa^2}{1-\varepsilon}}, \tag{34}$$

$$n_t = \frac{-2\varepsilon}{1-\varepsilon}. \tag{35}$$

Throughout this work, we simplify many elliptic integrals into an analytically tractable form; Ref. 22 contains tables of elliptic integrals, whereas Refs. 23 and 24 outline the method to reduce and simplify these integrals.

### A. Trapped particles

The longitudinal invariant for trapped particles is

$$J_b = \frac{8m\omega_0 l_0^2}{2\pi}\int_0^{\theta_b} d\theta\sqrt{\frac{\kappa^2-\sin^2(\theta/2)}{1+\varepsilon\cos(\theta)}}. \tag{36}$$

We perform the substitution

$$\theta = 2\,\arcsin\left(\sqrt{u}\kappa\right), \tag{37}$$

and find that

$$J_b = \frac{8m\omega_0 l_0^2}{2\pi\sqrt{2\varepsilon}}\int_0^1 du\frac{\sqrt{1-u}}{\sqrt{u}\sqrt{\frac{1}{\kappa^2}-u}\sqrt{\frac{1+\varepsilon}{2\varepsilon\kappa^2}-u}}. \tag{38}$$

Consulting Refs. 22–24, we find that this integral can be expressed analytically

$$J_b = \frac{4m\omega_0 l_0^2}{\pi\varepsilon}\sqrt{1+\varepsilon-2\varepsilon\kappa^2}\left(\frac{(1+\varepsilon)\Pi(n_b,k_b)}{1+\varepsilon-2\varepsilon\kappa^2}-K(k_b)\right). \tag{39}$$

In the small $\varepsilon$ limit, we find that

$$J_b \approx \frac{8m\omega_0 l_0^2}{\pi}\left(E(\kappa) - (1-\kappa^2)K(\kappa)\right). \tag{40}$$

Meanwhile, for small $\kappa$ we can approximate the bounce action as

$$J_b \approx \frac{2m\omega_0 l_0^2}{\sqrt{1+\varepsilon}}\kappa^2. \tag{41}$$

For both small $\kappa$ and small $\varepsilon$, it is then clear that

$$J_b \approx 2m\omega_0 l_0^2\kappa^2. \tag{42}$$

Next, we compute the period

$$T_b = \frac{4}{2\omega_0}\int_0^{\theta_b} d\theta\sqrt{\frac{1+\varepsilon\cos(\theta)}{\kappa^2 - \sin^2(\theta/2)}}. \tag{43}$$

We again perform the substitution $\theta = 2\arcsin\left(\sqrt{u}\kappa\right)$ and obtain

$$T_b = \frac{2\sqrt{2\varepsilon}}{\omega_0}\int_0^1 du\frac{\sqrt{\frac{1+\varepsilon}{2\varepsilon\kappa^2} - u}}{\sqrt{u}\sqrt{1-u}\sqrt{\frac{1}{\kappa^2} - u}}, \tag{44}$$

and using Refs. 22–24 once again,

$$T_b = \frac{4}{\omega_0}\frac{(1+\varepsilon)\Pi(n_b, k_b)}{\sqrt{1+\varepsilon - 2\varepsilon\kappa^2}}. \tag{45}$$

Therefore, the bounce frequency is

$$\omega_b = \frac{2\pi}{T_b} = \frac{\pi\omega_0}{2}\frac{\sqrt{1+\varepsilon-2\varepsilon\kappa^2}}{(1+\varepsilon)\Pi(n_b, k_b)}. \tag{46}$$

Taylor expanding the bounce period for small $\varepsilon$, we find that

$$T_b \approx \frac{4K(\kappa)}{\omega_0}, \tag{47}$$

so the bounce frequency can be approximated by

$$\omega_b \approx \frac{\pi\omega_0}{2K(\kappa)}. \tag{48}$$

In the limit of small $\kappa$, we instead have

$$T_b \approx \frac{2\pi}{\omega_0\sqrt{1+\varepsilon}}, \tag{49}$$

$$\omega_b \approx \omega_0\sqrt{1+\varepsilon}. \tag{50}$$

In the limit such that both $\varepsilon$ and $\kappa$ are small,

$$T_b \approx \frac{2\pi}{\omega_0}, \tag{51}$$

$$\omega_b \approx \omega_0. \tag{52}$$

## B. Passing particles

The longitudinal invariant for trapped particles is

$$J_t = \frac{4m\omega_0 l_0^2}{2\pi}\int_0^{\pi} d\theta\sqrt{\frac{\kappa^2 - \sin^2(\theta/2)}{1+\varepsilon\cos(\theta)}}. \tag{53}$$

We perform the substitution

$$\theta = 2\arcsin\left(\sqrt{u}\right), \tag{54}$$

which is similar to the substitution performed for passing particles. Upon doing so, we find that

$$J_t = \frac{4m\omega_0 l_0^2}{2\pi\sqrt{2\varepsilon}}\int_0^1 du\frac{\sqrt{\kappa^2 - u}}{\sqrt{u}\sqrt{1-u}\sqrt{\frac{1+\varepsilon}{2\varepsilon} - u}}. \tag{55}$$

Consulting Refs. 22–24, we find that this integral can be written analytically as

$$J_t = \frac{2m\omega_0 l_0^2}{\pi\varepsilon}\left(\frac{1+\varepsilon-2\varepsilon\kappa^2}{\kappa\sqrt{1-\varepsilon}}\right)\left(\frac{(1+\varepsilon)\Pi(n_t, k_t)}{1+\varepsilon-2\varepsilon\kappa^2} - K(k_t)\right). \tag{56}$$

Taking a Taylor expansion of $J_t$ about $\varepsilon = 0$, we obtain

$$J_t \approx \frac{4m\omega_0 l_0^2}{\pi}\kappa E(\kappa^{-1}). \tag{57}$$

Meanwhile, for large $\kappa$ transit action is approximately

$$J_t \approx \frac{m\omega_0 l_0^2\sqrt{2\varepsilon}}{\varepsilon(1+\varepsilon)}. \tag{58}$$

For both small $\varepsilon$ and large $\kappa$, the expression simplifies to

$$J_t \approx \sqrt{2mE}\bar{q}R_0 \approx m|v_\parallel|\bar{q}R_0. \tag{59}$$

The transit period is

$$T_t = \frac{2}{2\omega_0}\int_0^{\pi} d\theta\sqrt{\frac{1+\varepsilon\cos(\theta)}{\kappa^2 - \sin^2(\theta/2)}}. \tag{60}$$

Again performing the substitution $\theta = 2\arcsin\left(\sqrt{u}\right)$, we obtain

$$T_t = \frac{\sqrt{2\varepsilon}}{\omega_0}\int_0^1 du\frac{\sqrt{\frac{1+\varepsilon}{2\varepsilon} - u}}{\sqrt{u}\sqrt{1-u}\sqrt{\kappa^2 - u}}. \tag{61}$$

Using Refs. 22–24 again, we find that

$$T_t = \frac{2}{\omega_0}\frac{(1+\varepsilon)\Pi(n_t, k_t)}{\kappa\sqrt{1-\varepsilon}}. \tag{62}$$

Therefore, the transit frequency is

$$\omega_t = \frac{2\pi}{T_b} = \pi\omega_0\frac{\kappa\sqrt{1-\varepsilon}}{(1+\varepsilon)\Pi(n_t, k_t)}. \tag{63}$$

Taylor expanding the transit period for small $\varepsilon$, we find that

$$T_t \approx \frac{2K(\kappa^{-1})}{\omega_0\kappa}. \tag{64}$$

Therefore, the transit frequency can be approximated by

$$\omega_t \approx \frac{\pi\omega_0\kappa}{K(\kappa^{-1})}. \tag{65}$$

In the limit of large $\kappa$, we instead have

$$T_t \approx \frac{\pi\sqrt{2\varepsilon}}{\omega_0} = \frac{2\pi l_0}{\sqrt{2E/m}}, \tag{66}$$

$$\omega_t \approx \sqrt{\frac{2}{\varepsilon}}\omega_0 = \frac{\sqrt{2E/m}}{l_0}. \tag{67}$$

Finally, for small $\varepsilon$ and large $\kappa$, we obtain

$$T_t \approx \frac{2\pi l_0}{\sqrt{2E/m}} \approx \frac{2\pi \bar{q} R_0}{|v_\parallel|}, \tag{68}$$

$$\omega_t \approx \frac{\sqrt{2E/m}}{l_0} \approx \frac{|v_\parallel|}{\bar{q} R_0}. \tag{69}$$

### C. Comparison to the small $\varepsilon$ limit

Having derived analytical expressions for both the longitudinal invariant and the bounce-transit frequency, we can easily compare the exact expressions to the small $\varepsilon$ approximation. Figure 1 plots the longitudinal invariant while Fig. 2 plots the bounce-transit frequency for various values of $\varepsilon$, where $\varepsilon = 0$ refers to the small $\varepsilon$ limit. The $\varepsilon = 0.3$ case corresponds to the edge region in many tokamaks; $\varepsilon = 0.5, 0.7$ are also plotted to account for low aspect ratio tokamaks. We can see that the small $\varepsilon$ approximation is suitable even for the extreme case of $\varepsilon = 0.7$.

We note that the discontinuity in Fig. 1 at $\kappa = 1$ arises from the discrepancy in defining $J$ for the bounce and transit cases. $J$ is calculated with a closed line integral; a trapped particle undergoes a back and forth motion, whereas a passing particle's velocity never changes direction. Thus, there is a factor of 2 difference for the closed line integral depending on whether we consider the particle trapped or passing at $\kappa = 1$. No discontinuity arises in Fig. 2 since the frequency at the trapped-passing boundary is 0.

Having characterized the periodicity of the motion, we next develop closed form solutions to the equation of motion and find explicit expressions for $\theta(t)$.

## IV. CLOSED FORM SOLUTION TO THE BOUNCE-TRANSIT MOTION

To calculate the closed form solution for the bounce-transit motion, we reformulate the problem as follows: find $\theta(\alpha)$ where $\alpha(t)$ satisfies

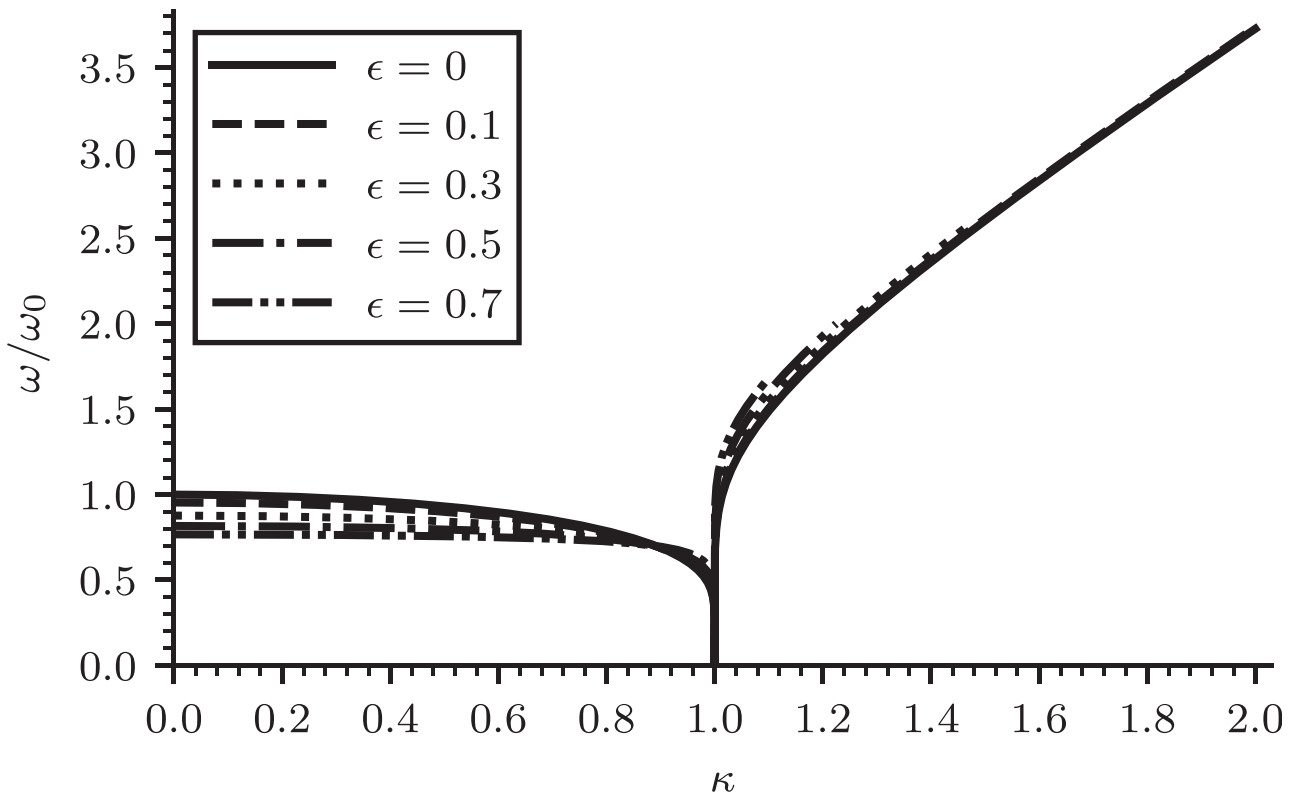

**FIG. 2.** Normalized bounce-transit frequency $\omega/\omega_0$ vs the trapping parameter $\kappa$. The bounce frequency corresponds to $\kappa < 1$ and the transit frequency corresponds to $\kappa > 1$. The upper bound for $\kappa$ is $\varepsilon$ dependent, so each case is plotted up until its respective upper bound.

$$\frac{d\alpha}{dt} = \omega. \tag{70}$$

The function $\alpha$ is called the action angle and is linked to the adiabatic invariant associated with the periodic motion. The strategy is to first find $\alpha(\theta)$ and then invert the function to obtain the closed form solution. Using the chain rule, we can write

$$\frac{d\alpha}{dt} = \frac{d\alpha}{d\theta}\dot{\theta} = \omega, \tag{71}$$

$$\alpha = \omega \int^{\theta} \frac{d\theta'}{\dot{\theta}'}. \tag{72}$$

The specific form of this integral is elliptic and depends on whether we consider trapped or passing particles. In the following calculations, we constrain $\alpha$ such that $-\pi \le \alpha \le \pi$ for one whole periodic motion and use the initial condition $\alpha(\theta = 0) = 0$. To invert the elliptic integral, we define a function $\mathrm{am}_\pi$ such that

$$\mathrm{am}_\pi(\Pi(\varphi, n, k), n, k) = \varphi. \tag{73}$$

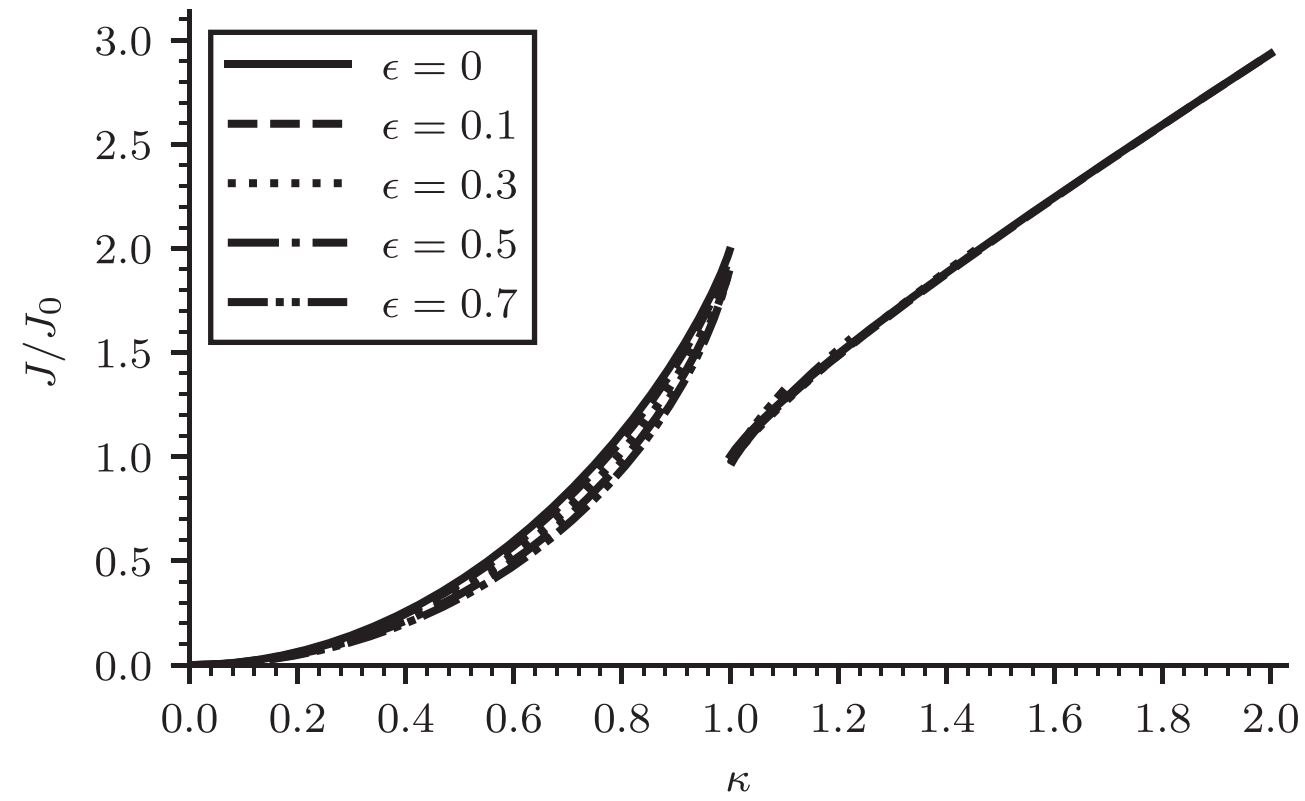

**FIG. 1.** Normalized longitudinal invariant $J/J_0$ vs the trapping parameter $\kappa$, where $J_0 = 8m\omega_0 l_0^2/\pi$. The bounce invariant corresponds to $\kappa < 1$ and the transit invariant corresponds to $\kappa > 1$. The upper bound for $\kappa$ is $\varepsilon$ dependent, so each case is plotted up until its respective upper bound.

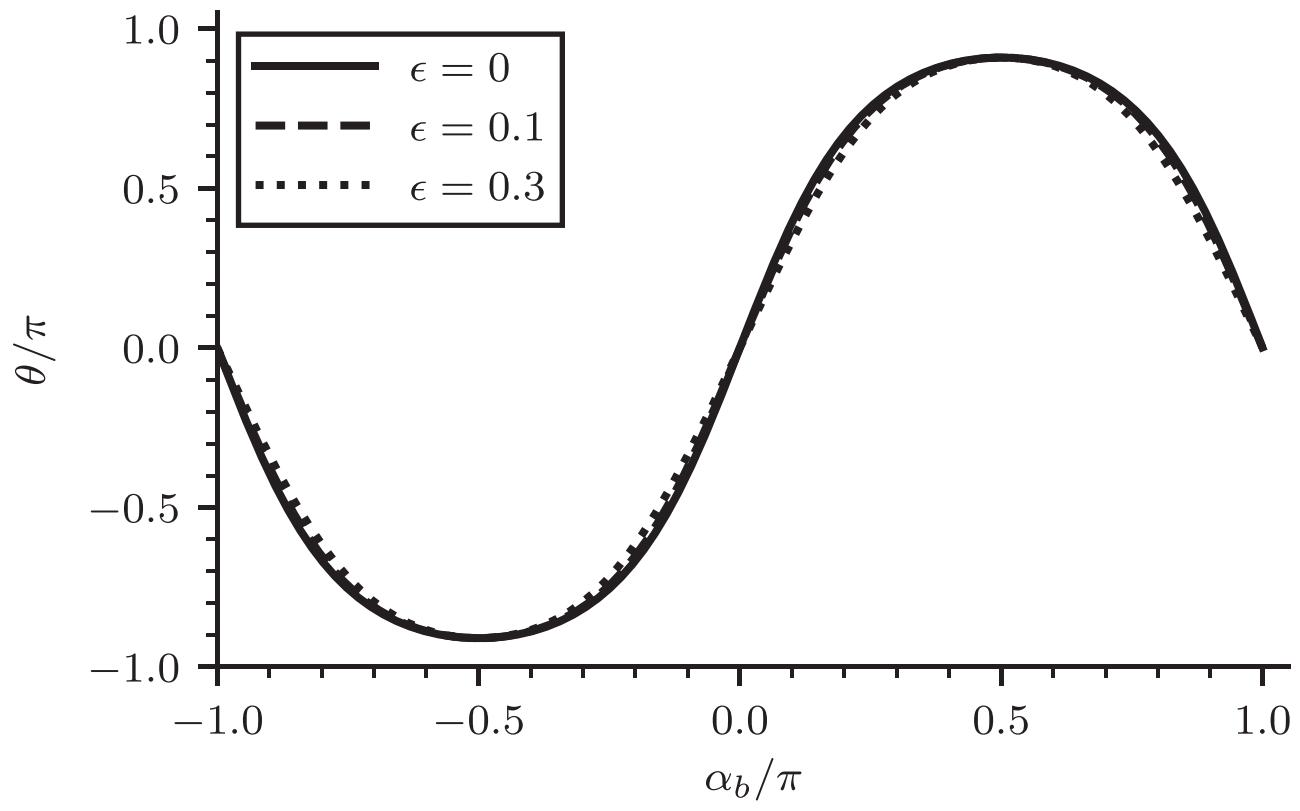

**FIG. 3.** Normalized poloidal angle $\theta/\pi$ vs normalized action angle $\alpha_b/\pi$ for $\kappa = 0.99$. The particle undergoes a nearly sinusoidal motion with little discrepancy between different values of $\varepsilon$.

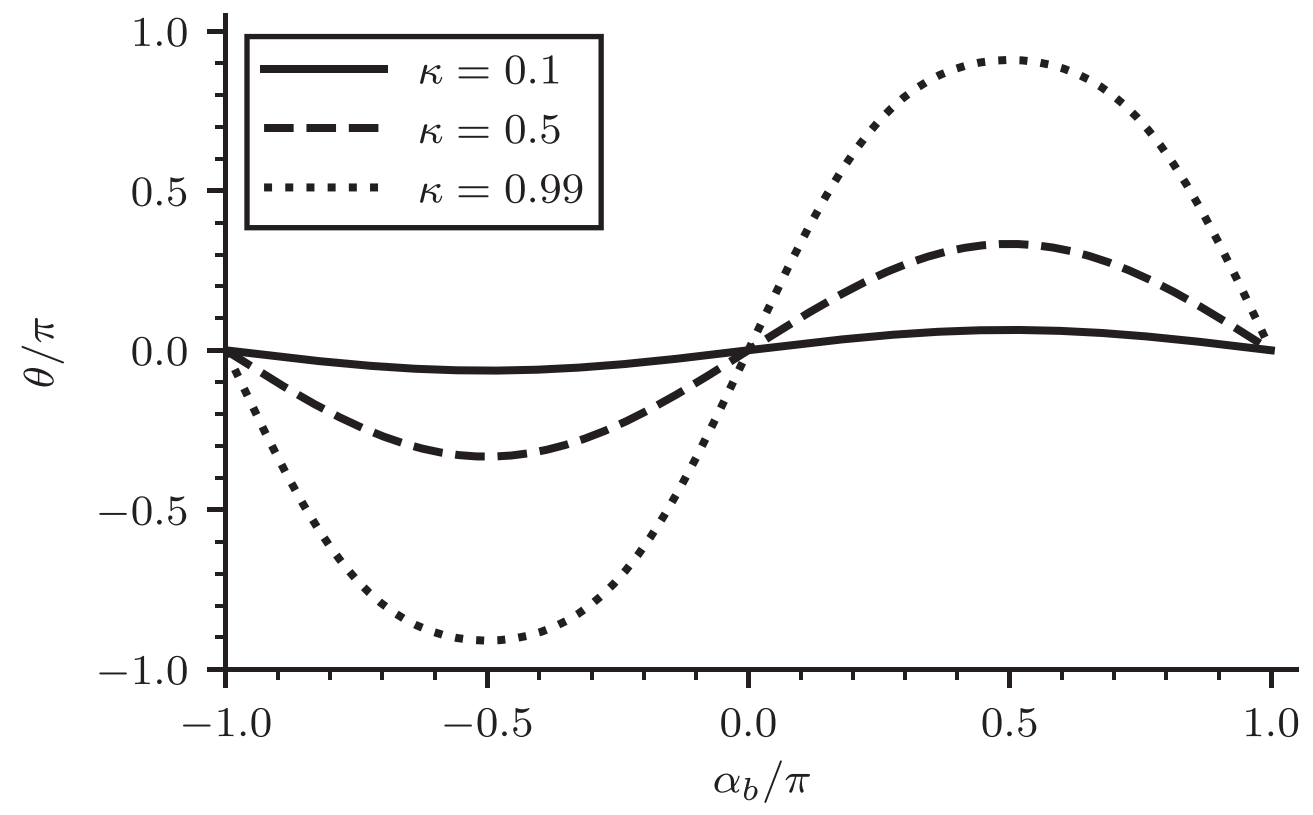

**FIG. 4.** Normalized poloidal angle $\theta/\pi$ vs normalized action angle $\alpha_b/\pi$ for $\varepsilon = 0.3$. The amplitude of the motion increases with $\kappa$ as expected.

This is analogous to the Jacobi amplitude function that is used for elliptic integrals of the first kind. For ease of notation, we drop $n$ and $k$. We also define analogous trigonometric functions,

$$\mathrm{sn}_\pi(u) = \sin(\mathrm{am}_\pi(u)), \tag{74}$$

$$\mathrm{cn}_\pi(u) = \cos(\mathrm{am}_\pi(u)), \tag{75}$$

$$\mathrm{dn}_\pi(u) = \sqrt{1 - k^2\,\mathrm{sn}_\pi(u)}\left(1 - \alpha^2\,\mathrm{sn}_\pi(u)\right). \tag{76}$$

These functions satisfy the relations

$$\mathrm{sn}_\pi^2(u) + \mathrm{cn}_\pi^2(u) = 1, \tag{77}$$

$$\frac{d}{du}\mathrm{sn}_\pi(u) = \mathrm{cn}_\pi(u)\,\mathrm{dn}_\pi(u), \tag{78}$$

$$\frac{d}{du}\mathrm{cn}_\pi(u) = -\,\mathrm{sn}_\pi(u)\,\mathrm{dn}_\pi(u). \tag{79}$$

We clarify that these functions do not have the same properties as the typical Jacobi elliptic functions, so care must be taken in using them.

## A. Trapped particles

For trapped particles, the integral we wish to compute can be simplified to

$$\alpha_b = \frac{\pi}{2}\frac{\displaystyle\int_0^\theta d\theta'\sqrt{\frac{1+\varepsilon\cos(\theta')}{\kappa^2 - \sin^2(\theta'/2)}}}{\displaystyle\int_0^{\theta_b} d\theta'\sqrt{\frac{1+\varepsilon\cos(\theta')}{\kappa^2 - \sin^2(\theta'/2)}}}, \tag{80}$$

where for convenience we temporarily restrict our domain to $0 \le \alpha_b \le \pi/2$ and $0 \le \theta \le \theta_b$. We define a new angle $\varphi_b$

$$\sin^2(\varphi_b) = \frac{\sin^2(\theta/2)}{\kappa^2}\frac{1+\varepsilon - 2\varepsilon\kappa^2}{1+\varepsilon - 2\varepsilon\kappa^2\sin^2(\theta/2)} \tag{81}$$

and obtain

$$\alpha_b = \frac{\pi}{2}\frac{\Pi(\varphi_b, n_b, k_b)}{\Pi(n_b, k_b)}. \tag{82}$$

As a sanity check, we note that $\varphi = \pi/2$ corresponds to $\theta = \theta_b$, in which case the numerator (having become a complete integral) cancels with the denominator. Using our previously defined inverse functions, we find that

$$\sin^2(\varphi_b) = \mathrm{sn}_\pi^2\left(\frac{2\alpha_b}{\pi}\Pi(n_b, k_b), n_b, k_b\right). \tag{83}$$

We abbreviate the right-hand side as $\mathrm{sn}_\pi^2\left(\frac{2\alpha_b}{\pi}\Pi\right)$. We can then solve for $\theta$,

$$\sin(\theta/2) = \kappa\,\mathrm{sn}_\pi\left(\frac{2\alpha_b}{\pi}\Pi\right)\sqrt{\frac{1+\varepsilon}{1+\varepsilon - 2\varepsilon\kappa^2\,\mathrm{cn}_\pi^2\left(\frac{2\alpha_b}{\pi}\Pi\right)}}, \tag{84}$$

$$\cos(\theta/2) = \sqrt{\frac{1+\varepsilon - 2\varepsilon\kappa^2 - \kappa^2(1-\varepsilon)\,\mathrm{sn}_\pi^2\left(\frac{2\alpha_b}{\pi}\Pi\right)}{1+\varepsilon - 2\varepsilon\kappa^2\,\mathrm{cn}_\pi^2\left(\frac{2\alpha_b}{\pi}\Pi\right)}}. \tag{85}$$

These equations are now valid for all $\alpha_b$ and by extension valid for all time. Taking the time derivative, we find that

$$\dot\theta = \frac{2\kappa\omega_0\,\mathrm{cn}_\pi\left(\frac{2\alpha_b}{\pi}\Pi\right)}{\sqrt{1+\varepsilon}}. \tag{86}$$

Thus, we have closed form expressions for $\theta$ and $\dot\theta$ as functions of $\alpha_b(t)$.

In the small $\varepsilon$ limit, we can approximate these expressions as

$$\sin(\theta/2) \approx \kappa\,\mathrm{sn}\left(\frac{2\alpha_b}{\pi}K(\kappa), \kappa\right), \tag{87}$$

$$\cos(\theta/2) \approx \kappa\,\mathrm{cn}\left(\frac{2\alpha_b}{\pi}K(\kappa), \kappa\right), \tag{88}$$

$$\dot\theta \approx 2\kappa\omega_0\,\mathrm{cn}\left(\frac{2\alpha_b}{\pi}K(\kappa), \kappa\right), \tag{89}$$

where sn and cn are the usual Jacobi elliptic functions. $\theta$ and $\dot\theta$ for trapped particles are plotted in Figs. 3–6.

## B. Passing particles

For passing particles, we instead calculate

$$\alpha_t = \pi\frac{\displaystyle\int_0^\theta d\theta'\sqrt{\frac{1+\varepsilon\cos(\theta')}{\kappa^2 - \sin^2(\theta'/2)}}}{\displaystyle\int_0^{\pi} d\theta'\sqrt{\frac{1+\varepsilon\cos(\theta')}{\kappa^2 - \sin^2(\theta'/2)}}}. \tag{90}$$

Similar to $\varphi_b$, we define a new variable $\varphi_t$

$$\sin^2(\varphi_t) = \sin^2(\theta/2)\frac{1-\varepsilon}{1+\varepsilon - 2\varepsilon\sin^2(\theta/2)} \tag{91}$$

and obtain

$$\alpha_t = \pi\frac{\Pi(\varphi_t, n_t, k_t)}{\Pi(n_t, k_t)}. \tag{92}$$

Note that $\varphi = \pi/2$ corresponds to $\theta = \pi$. Performing the inversion, we find that

$$\sin^2(\varphi_t) = \mathrm{sn}_\pi^2\left(\frac{\alpha_t}{\pi}\Pi(n_t, k_t), n_t, k_t\right). \tag{93}$$

Abbreviating the right-hand side as $\mathrm{sn}_\pi^2(\frac{\alpha_t}{\pi}\Pi)$, we can then solve for $\theta$,

$$\sin(\theta/2) = \mathrm{sn}_\pi\left(\frac{\alpha_t}{\pi}\Pi\right)\sqrt{\frac{1+\varepsilon}{1+\varepsilon-2\varepsilon\,\mathrm{cn}_\pi^2\left(\frac{\alpha_t}{\pi}\Pi\right)}}, \tag{94}$$

$$\cos(\theta/2) = \mathrm{cn}_\pi\left(\frac{\alpha_t}{\pi}\Pi\right)\sqrt{\frac{1-\varepsilon}{1+\varepsilon-2\varepsilon\,\mathrm{cn}_\pi^2\left(\frac{\alpha_t}{\pi}\Pi\right)}}. \tag{95}$$

These equations are valid for all $\alpha_t$ and thus all time.

We can also calculate $\dot{\theta}$ for passing particles. However, we need to take into account the fact that the parallel velocity can either be positive or negative and never changes sign. To accommodate for the potential sign difference, we define $\varepsilon_\parallel$ to be the sign of $v_\parallel$

$$v_\parallel = \varepsilon_\parallel |v_\parallel|. \tag{96}$$

Then, since we previously defined $\omega_t$ to be strictly positive, we need to write

$$\frac{d\alpha_t}{dt} = \varepsilon_\parallel \omega_t. \tag{97}$$

We can then proceed to safely calculating $\dot{\theta}$ by taking a time derivative and simplifying

$$\dot{\theta} = \frac{2\kappa\omega_0\varepsilon_\parallel\sqrt{1-\frac{1+\varepsilon-2\varepsilon\kappa^2}{\kappa^2(1-\varepsilon)}\,\mathrm{sn}_\pi^2\left(\frac{\alpha_t}{\pi}\Pi\right)}}{\sqrt{1+\varepsilon}}. \tag{98}$$

We have now obtained closed form solutions for $\theta$ and $\dot{\theta}$ that are valid for all time.

In the small $\varepsilon$ limit, these expressions are approximately

$$\sin(\theta/2) \approx \mathrm{sn}\left(\frac{\alpha_t}{\pi}K(\kappa^{-1}), \kappa^{-1}\right), \tag{99}$$

$$\cos(\theta/2) \approx \mathrm{cn}\left(\frac{\alpha_t}{\pi}K(\kappa^{-1}), \kappa^{-1}\right), \tag{100}$$

$$\dot{\theta} \approx 2\kappa\omega_0\varepsilon_\parallel\,\mathrm{dn}\left(\frac{\alpha_t}{\pi}K(\kappa^{-1}), \kappa^{-1}\right). \tag{101}$$

$\theta$ and $\dot{\theta}$ for passing particles are plotted in Figs. 7–10.

### C. Phase space portrait

Having calculated the closed form solutions for the bounce-transit motion, we can easily plot the phase-space portrait as shown in Fig. 5 for $\varepsilon = 0.3$ and Fig. 6 for $\varepsilon = 0$. The bounce orbits reside in the interior of the separatrix where the particle follows a closed loop. On the exterior of the separatrix, the particle instead continues its motion such that $\dot{\theta}$ never changes sign. The phase-space portrait for the case of $\varepsilon = 0$ is mathematically identical to that of a simple pendulum. While the portrait for $\varepsilon = 0.3$ looks qualitatively similar to a simple pendulum's, there is an important difference. In the case of a simple pendulum, the phase portrait is unbounded in the $\dot{\theta}$ direction; one can find an orbit corresponding to any value of $\dot{\theta}_{\max}/\omega_{0,\text{pend.}}$. For transit orbits in a concentric circular geometry, the phase-space orbit curves are bounded by the curve corresponding to $\kappa = \sqrt{(1+\varepsilon)/(2\varepsilon)}$. We can see in Fig. 11 that the outermost curves corresponding to $\kappa = 1.4$ are very straight and already nearly bound any interior curves. In contrast, for the small $\varepsilon$ limit $\kappa$ is unbounded from above, so a phase-space orbit curve can be found for any value of $\dot{\theta}_{\max}/\omega_0$. This can be seen in Fig. 12 where the outermost curves are not straight; indeed, they never become perfectly straight, regardless of what value $\kappa$ takes.

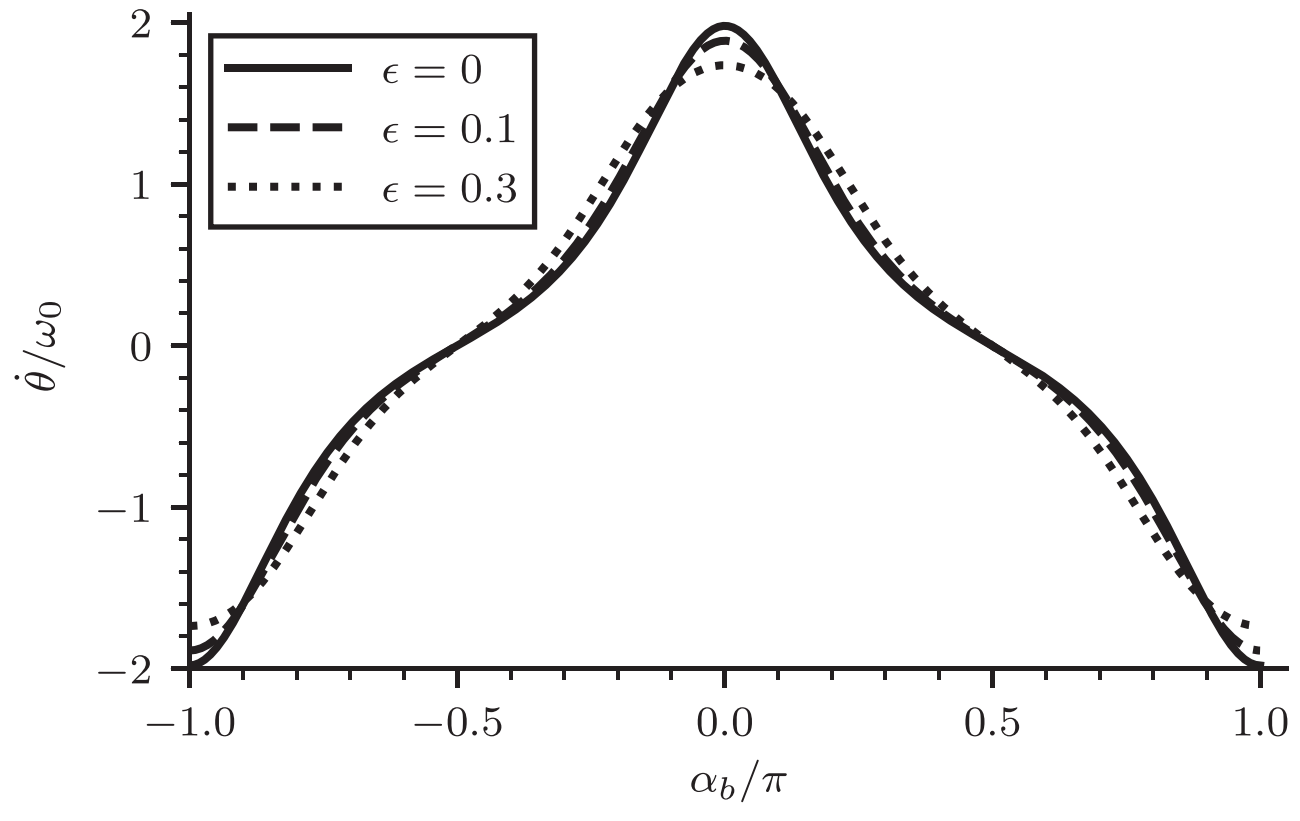

**FIG. 5.** Normalized poloidal velocity $\dot{\theta}/\omega_0$ vs the normalized action angle $\alpha_b/\pi$ for $\kappa = 0.99$. It is more clear that the particle's motion deviates from sinusoidal motion.

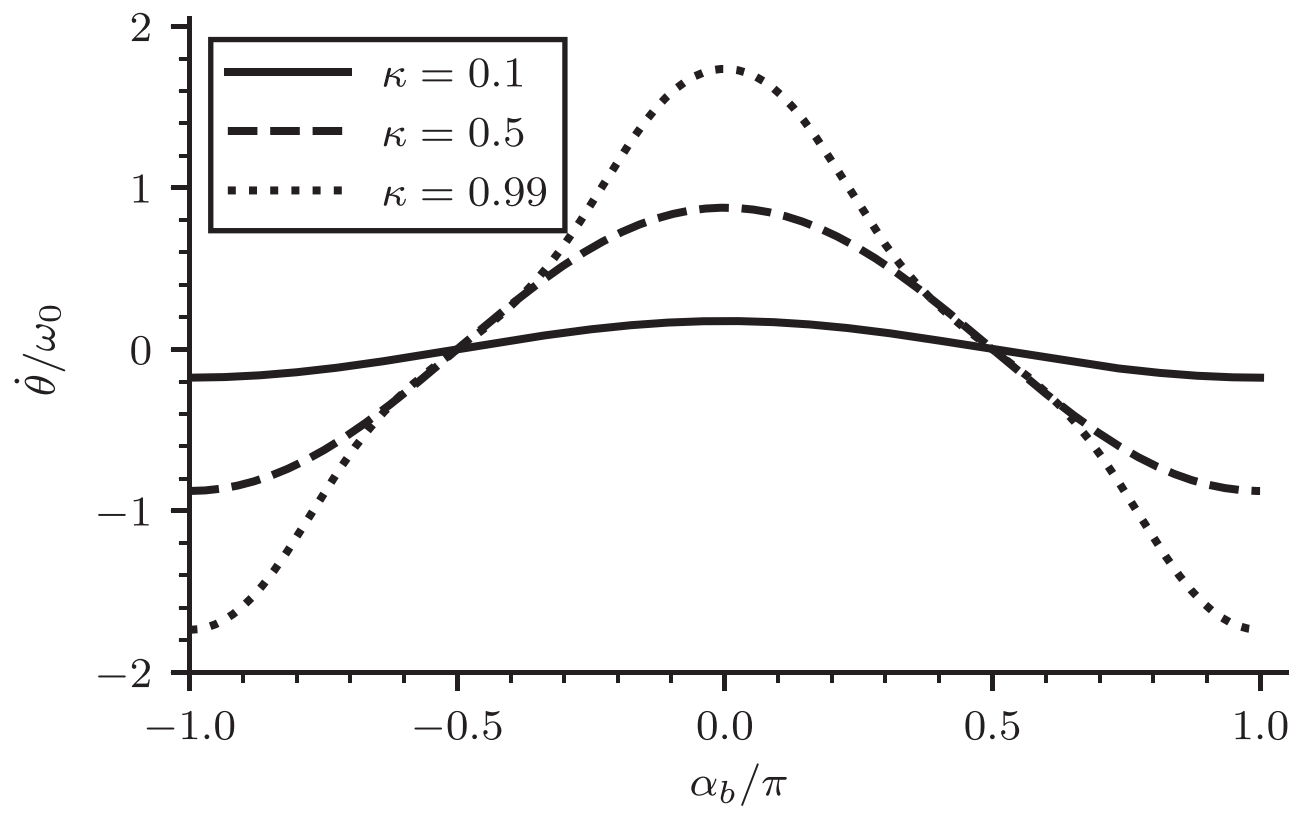

**FIG. 6.** Normalized poloidal velocity $\dot{\theta}/\omega_0$ vs the normalized action angle $\alpha_b/\pi$ for $\varepsilon = 0.3$. The amplitude of the motion increases with $\kappa$ as expected.

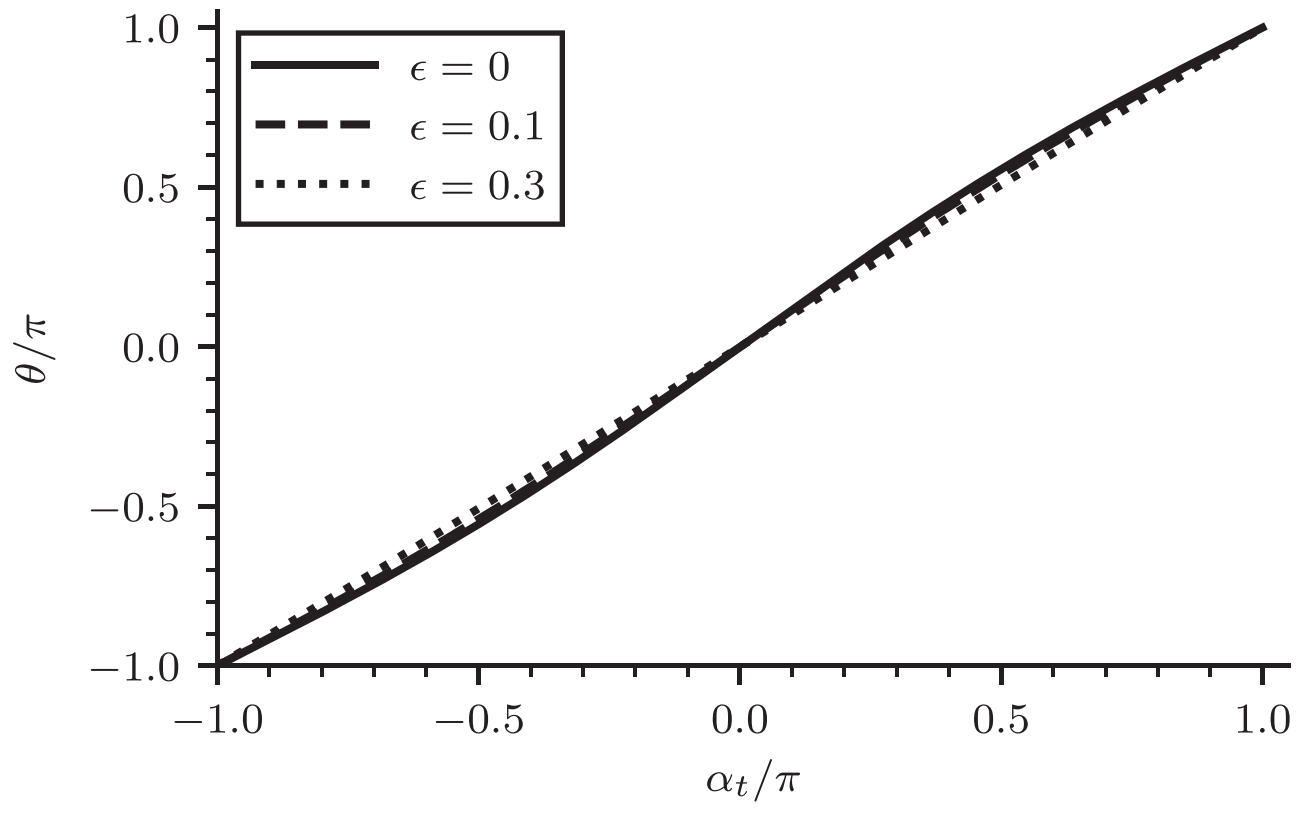

**FIG. 7.** Normalized poloidal angle $\theta/\pi$ vs normalized action angle $\alpha_t/\pi$ for $\kappa = 1.4$. In the extreme passing limit the particle has a nearly constant poloidal velocity, with very little deviation between different values of $\varepsilon$.

This completes our discussion of the lowest-order guiding center orbit. Without assuming anything about $\varepsilon$ in a concentric circular geometry, we have derived analytical expressions for the bounce-transit frequencies and closed form solutions to the bounce-transit equation of motion. We then proceed to the next order in the gyroradius by considering deviations from the field line.

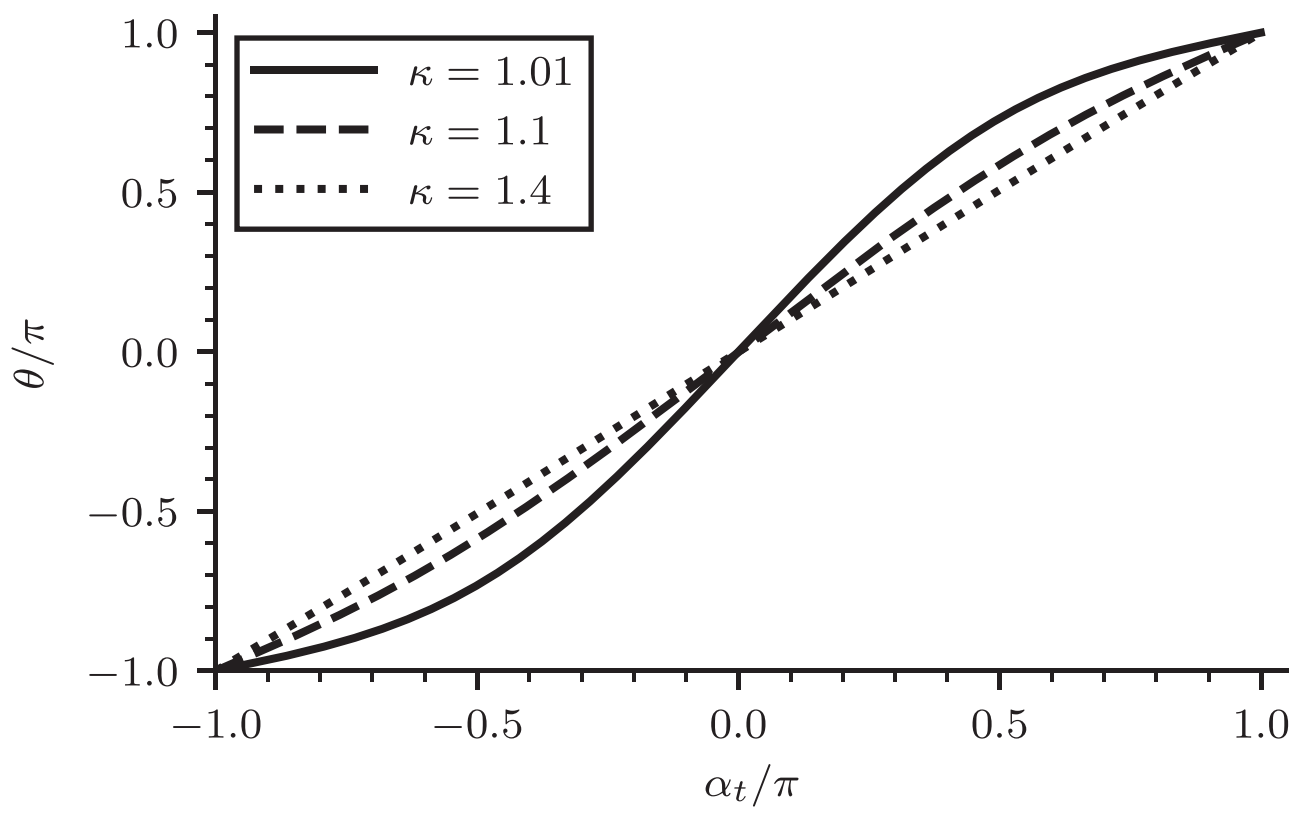

**FIG. 8.** Normalized poloidal angle $\theta/\pi$ vs normalized action angle $\alpha_t/\pi$ for $\varepsilon = 0.3$. As $\kappa$ increases, the motion becomes straighter and less sinusoidal.

## V. BANANA ORBITS

First, we consider the radial deviation from the lowest-order guiding center orbit. To calculate the radial deviation, we expand about a reference magnetic surface $\psi = \psi_0$, noting that $\psi = \psi(r)$. We define a reference radial position $r_0$ such that

$$\psi_0 = \psi(r_0), \tag{102}$$

$$P_\phi = -e\psi + \frac{mB_0R_0v_\parallel}{B} = -e\psi_0, \tag{103}$$

where $P_\phi$ is the canonical toroidal momentum. Performing an expansion about $r_0$, we find that

$$r \approx r_0 + \left.\frac{dr}{d\psi}\right|_{\psi_0}(\psi - \psi_0) = r_0 + \frac{\bar{q}R_0}{r_0}\frac{mv_\parallel}{eB}. \tag{104}$$

We can simplify by writing $v_\parallel$ and $B$ in terms of $\theta$ and $\dot{\theta}$, which in turn are functions of the action angle $\alpha$. For trapped particles, we obtain

$$r \approx r_0 + 2\frac{\bar{q}^2R_0^2}{r_0}\frac{m\omega_0}{eB_0}\frac{\kappa\sqrt{1+\varepsilon}(1+\varepsilon-2\varepsilon\kappa^2)\,\text{cn}_\pi\left(\frac{2\alpha_b}{\pi}\Pi\right)}{1+\varepsilon-2\varepsilon\kappa^2\,\text{cn}_\pi^2\left(\frac{2\alpha_b}{\pi}\Pi\right)}. \tag{105}$$

We isolate one of the factors and find that

$$\frac{\bar{q}^2R_0^2}{r_0}\frac{m\omega_0}{eB_0} = \frac{\bar{q}}{\Omega_c\sqrt{\varepsilon}}\sqrt{\frac{E}{m}}, \tag{106}$$

where $\Omega_c = eB^0/m$ is the characteristic cyclotron frequency. We then write

$$\frac{\bar{q}}{\Omega_c\sqrt{\varepsilon}}\sqrt{\frac{E}{m}} = \frac{\bar{q}\rho_c}{\sqrt{2\varepsilon}}\sqrt{\frac{E}{\mu B^0}} = \frac{\delta_b}{\sqrt{1+\varepsilon-2\varepsilon\kappa^2}}, \tag{107}$$

where we have used the cyclotron radius $\rho_c$ and the banana width

$$\delta_b = \frac{\bar{q}\rho_c}{\sqrt{2\varepsilon}}. \tag{108}$$

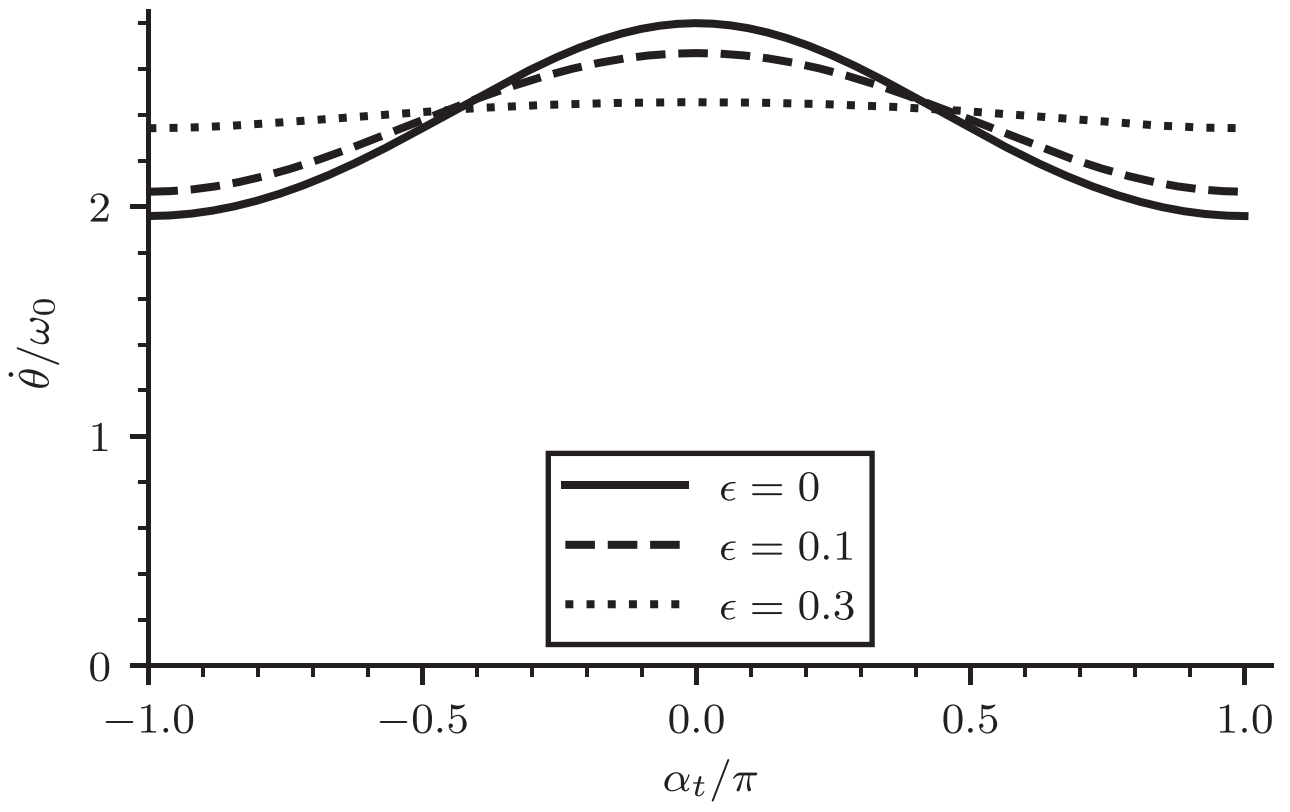

**FIG. 9.** Normalized poloidal velocity $\dot{\theta}/\omega_0$ the vs the normalized action angle $\alpha_t/\pi$ for $\kappa = 1.4$. $\kappa = 1.4$ is very close to the maximum allowed trapping parameter for $\varepsilon = 0.3$, hence why its plot is the straightest.

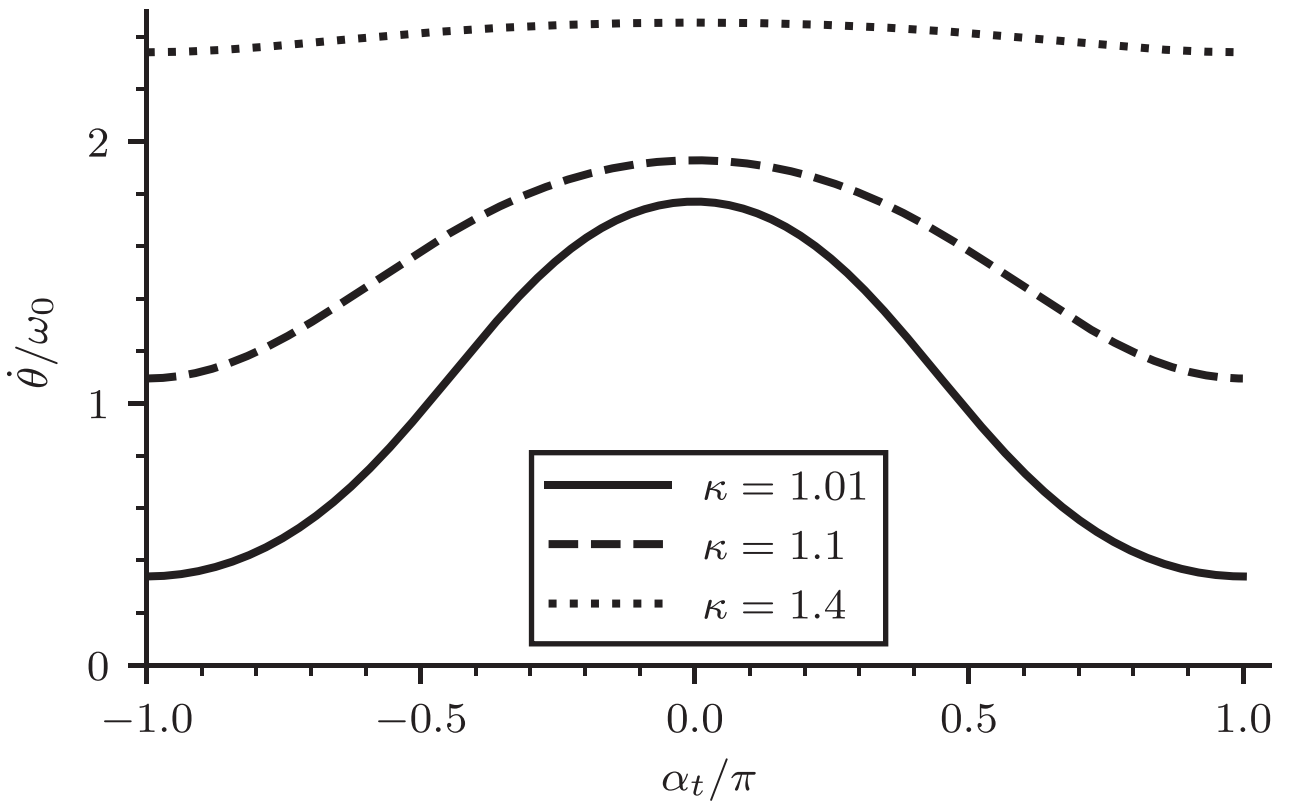

**FIG. 10.** Normalized poloidal velocity $\dot{\theta}/\omega_0$ vs the normalized action angle $\alpha_t/\pi$ for $\varepsilon = 0.3$. The maximum velocity increases for larger $\kappa$ and in turn the variation of the velocity decreases.

Therefore, for trapped particles, the expression simplifies to

$$r \approx r_0 + 2\delta_b\kappa \frac{\sqrt{1+\varepsilon}\sqrt{1+\varepsilon-2\varepsilon\kappa^2}\,\mathrm{cn}_\pi\left(\frac{2\alpha_b}{\pi}\Pi\right)}{1+\varepsilon-2\varepsilon\kappa^2\,\mathrm{cn}_\pi^2\left(\frac{2\alpha_b}{\pi}\Pi\right)}. \tag{109}$$

The banana orbits are plotted in the poloidal plane in Fig. 13, where the characteristic banana shape is readily apparent.

For passing particles, we instead obtain the expression

$$r \approx r_0 + 2\varepsilon_\parallel\delta_b\kappa\sqrt{\frac{1-\varepsilon^2}{1+\varepsilon-2\varepsilon\kappa^2}} \times \frac{\sqrt{1-\varepsilon-(1+\varepsilon-2\varepsilon\kappa^2)\,\mathrm{sn}_\pi^2\left(\frac{\alpha_t}{\pi}\Pi\right)}}{1+\varepsilon-2\varepsilon\,\mathrm{cn}_\pi^2\left(\frac{\alpha_t}{\pi}\Pi\right)}. \tag{110}$$

However, the radial deviation from $\psi_0$ can be rather large for passing particles given a high enough velocity. If we instead expand about the time average of $\psi$, we would obtain an expression with a lower deviation. It is easily shown that in our case

$$\langle\psi\rangle = \psi_0 + \frac{mR_0^2\bar{q}\varepsilon_\parallel\omega_t}{e}. \tag{111}$$

Defining $\langle r\rangle$ such that $\psi(\langle r\rangle) = \langle\psi\rangle$, we find that the equation for the radial deviation is modified in the following way:

$$r \approx \langle r\rangle + 2\varepsilon_\parallel\delta_b\kappa\sqrt{\frac{1-\varepsilon^2}{1+\varepsilon-2\varepsilon\kappa^2}} \times \frac{\sqrt{1-\varepsilon-(1+\varepsilon-2\varepsilon\kappa^2)\,\mathrm{sn}_\pi^2\left(\frac{\alpha_t}{\pi}\Pi\right)}}{1+\varepsilon-2\varepsilon\mathrm{cn}_\pi^2\left(\frac{\alpha_t}{\pi}\Pi\right)} - \frac{\pi\delta_b\varepsilon_\parallel\kappa\sqrt{1-\varepsilon}}{(1+\varepsilon)\Pi(n_t,k_t)\sqrt{1+\varepsilon-2\varepsilon\kappa^2}}. \tag{112}$$

We can compare the two methods by plotting the transit orbits in the poloidal plane as shown in Figs. 14 and 15. As predicted, expanding from the time averaged flux surface corresponds to less deviation from the flux surface.

Having studied the radial excursion from the magnetic field line, we now calculate the change in the toroidal motion after accounting for magnetic drifts and the conservation of canonical toroidal momentum.

## VI. TOROIDAL DRIFT FREQUENCY

Magnetic drifts and the conservation of canonical toroidal momentum alter the toroidal motion so that it no longer follows the magnetic field line. Rather, the guiding center drifts toroidally through the magnetic field lines lying on a flux surface. The toroidal drift frequency characterizes this motion and can be calculated by time averaging the equation of motion corresponding to the toroidal direction.

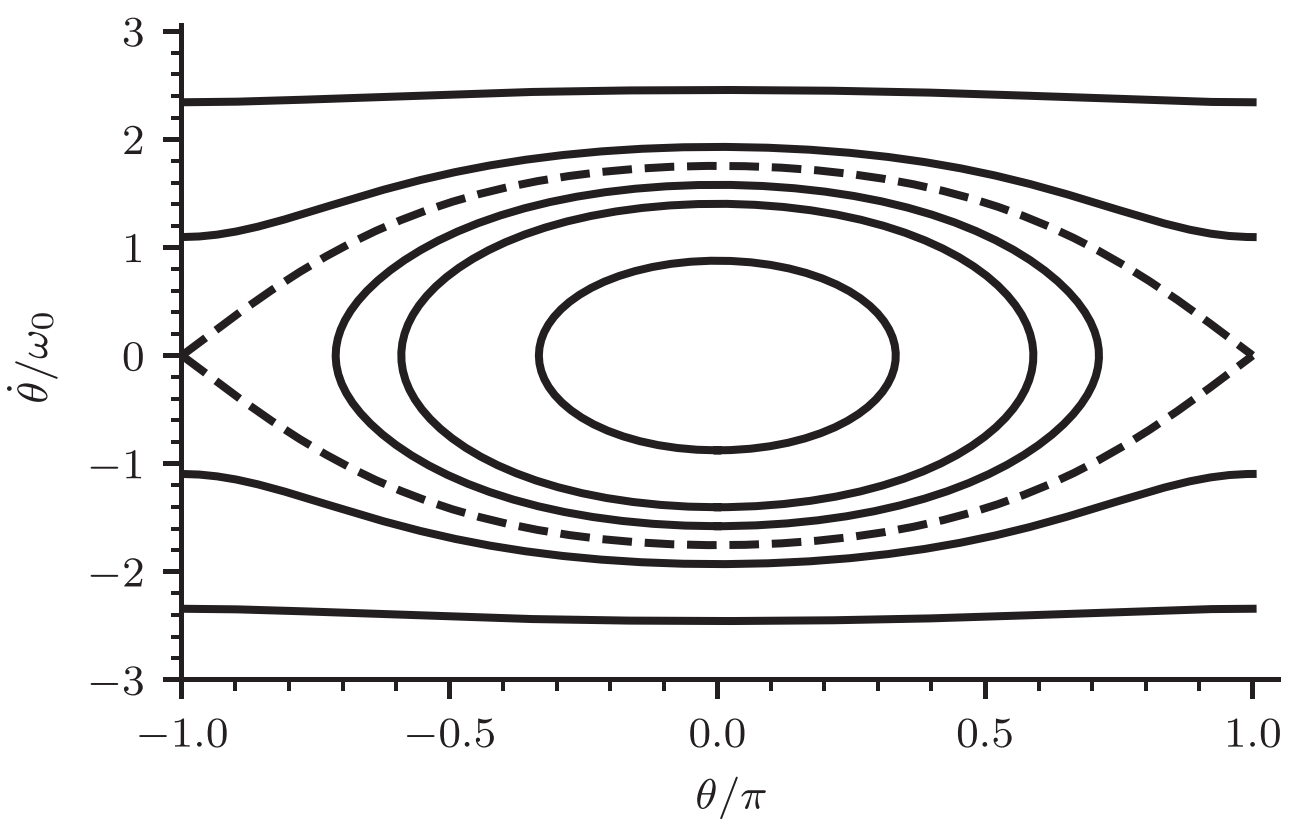

FIG. 11. Phase space plot of $\dot\theta/\omega_0$ vs $\theta/\pi$ for $\varepsilon = 0.3$. The separatrix represented by the dotted curve corresponds to the trapped-passing boundary at $\kappa = 1$.

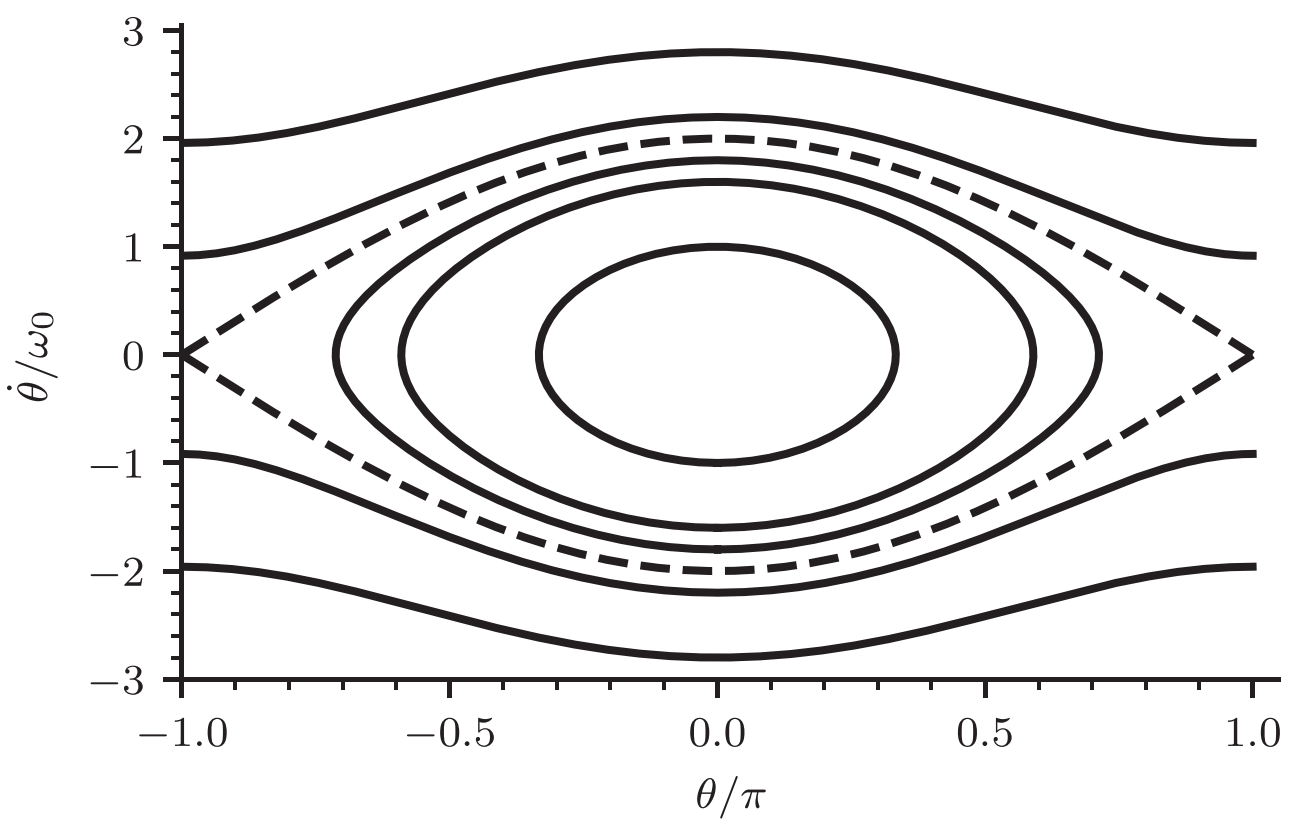

FIG. 12. Phase space plot of $\dot\theta/\omega_0$ vs $\theta/\pi$ for $\varepsilon = 0$. The separatrix represented by the dotted curve corresponds to the trapped-passing boundary at $\kappa = 1$.

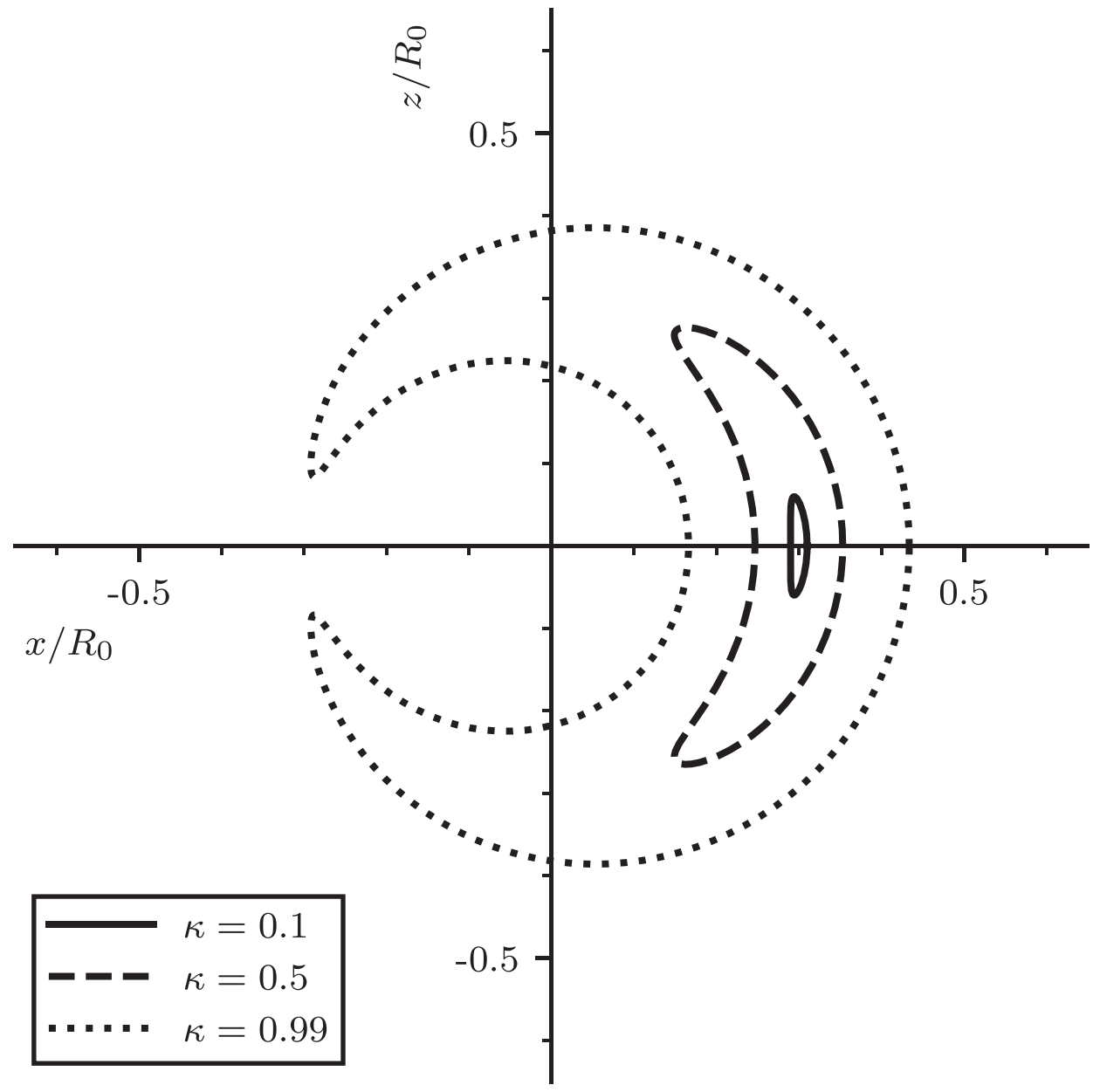

FIG. 13. Banana orbits in the normalized poloidal $(x/R_0, z/R_0)$ plane for $\varepsilon = 0.3$, $\delta_b/R_0 = 0.05$, $\psi_0$ as the reference magnetic flux surface. The origin corresponds to the magnetic axis.

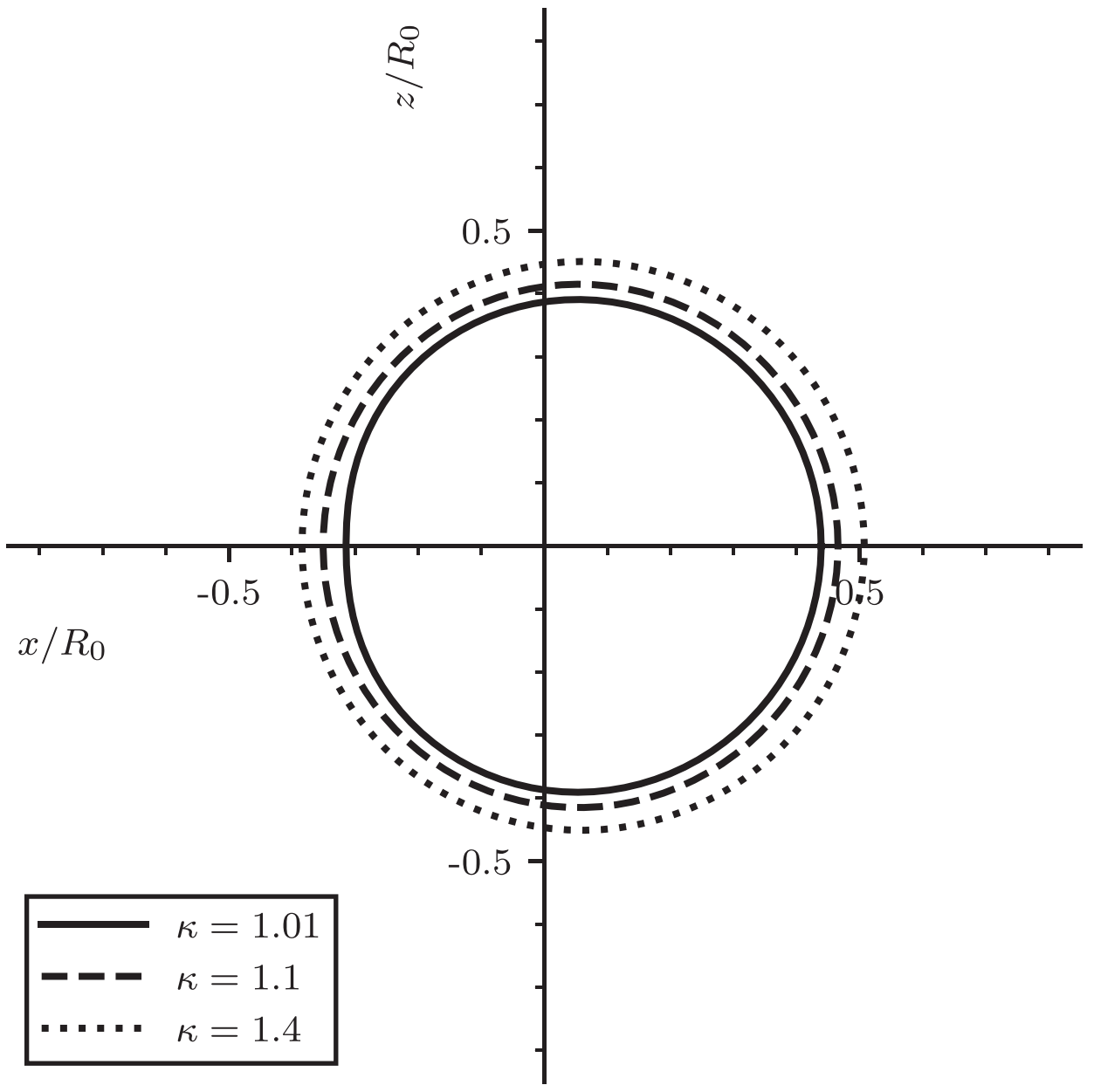

**FIG. 14.** Transit orbits in the normalized poloidal $(x/R_0, z/R_0)$ plane for $\varepsilon = 0.3$, $\delta_b/R_0 = 0.05$, using $\psi_0$ as the reference magnetic flux surface. The origin corresponds to the magnetic axis.

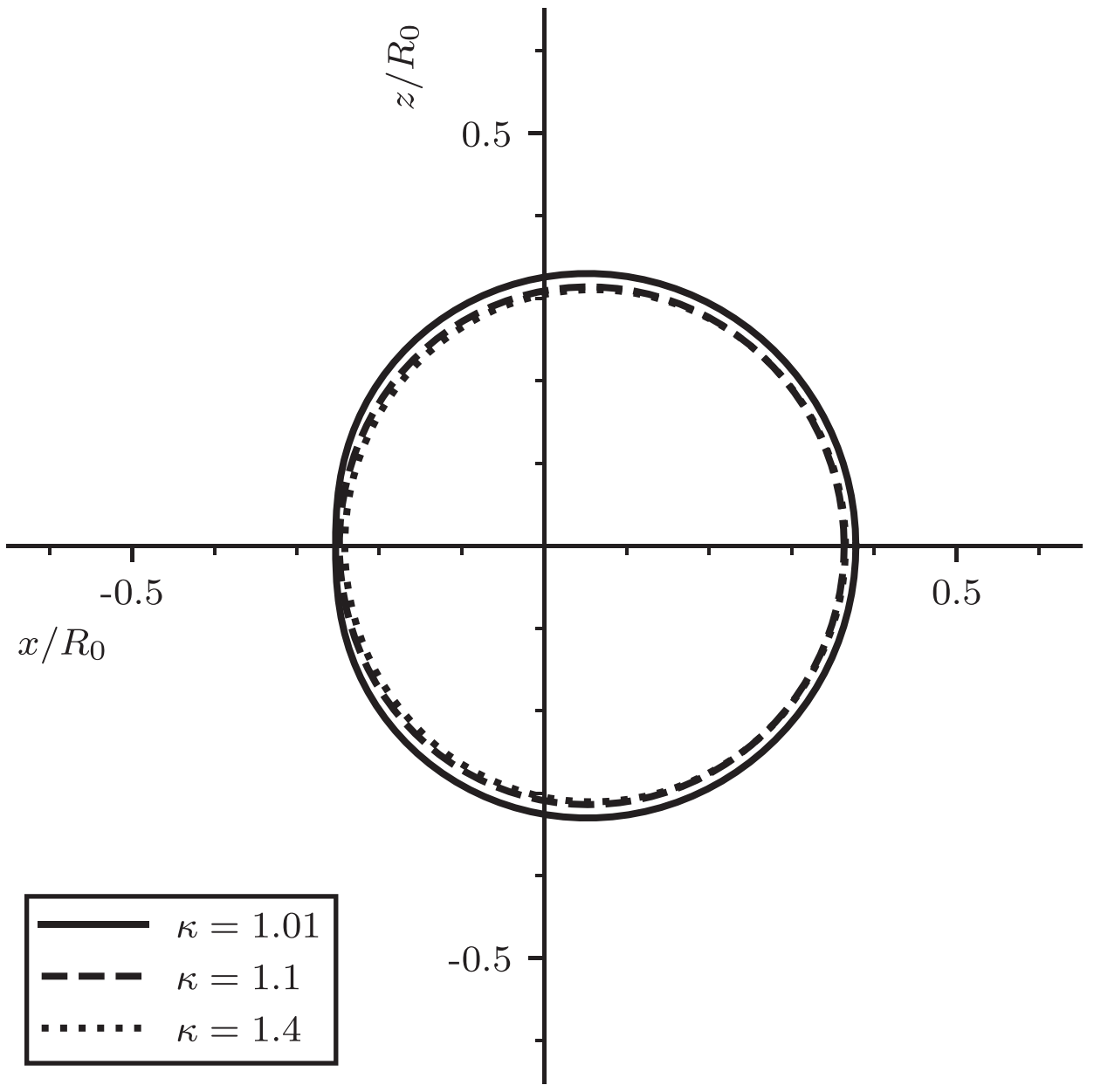

**FIG. 15.** Banana orbits in the normalized poloidal $(x/R_0,\ z/R_0)$ plane for $\varepsilon = 0.3$, $\delta_b/R_0 = 0.05$, using $\langle\psi\rangle$ as the reference magnetic flux surface. The origin corresponds to the magnetic axis. As expected, expanding about the time averaged flux surface results in a smaller deviation.

As in the case of banana orbits, we define a reference magnetic surface $\psi = \psi_0$ and Taylor expand around it; because the radial excursion is proportional to the gyroradius, this is equivalent to a gyroradius expansion. We note that the exactly conserved canonical toroidal momentum is again

$$P_\phi = -e\psi + \frac{mB_0R_0v_\parallel}{B} = -e\psi_0. \tag{113}$$

We then find that

$$\psi = \psi_0 + \frac{mB_0R_0v_\parallel}{eB} = \psi_0 + \psi_1. \tag{114}$$

The guiding center equations of motion without an electric field in a static magnetic field are given by

$$\frac{dx}{dt} = (v_\parallel\hat{\mathbf{b}} + \mathbf{v}_D)\cdot\nabla x, \tag{115}$$

$$\frac{dv_\parallel}{dt} = -\frac{\mu}{m}\nabla B, \tag{116}$$

where $\mathbf{v}_D$ is the magnetic drift. We apply these equations of motion to the coordinates $\psi, \chi, \phi$ and expand

$$\frac{d\psi}{dt} \approx \mathbf{v}_D\cdot\nabla\psi\,|_{\psi_0}, \tag{117}$$

$$\frac{d\chi}{dt} \approx \Omega_\chi\,|_{\psi_0} + \left.\frac{d\Omega_\chi}{d\psi}\right|_{\psi_0}\psi_1 + \mathbf{v}_D\cdot\nabla\chi\,|_{\psi_0}, \tag{118}$$

$$\frac{d\phi}{dt} \approx q\Omega_\chi\,|_{\psi_0} + \left.\frac{d}{d\psi}(q\Omega_\chi)\right|_{\psi_0}\psi_1 + \mathbf{v}_D\cdot\nabla\phi\,|_{\psi_0}, \tag{119}$$

where the lowest-order magnetic drift is the sum of the classical curvature and grad-$B$ drifts

$$\mathbf{v}_D = \frac{\hat{\mathbf{b}}}{eB}\times\left(mv_\parallel^2\mathbf{k} + \mu\nabla B\right). \tag{120}$$

Here, $e$ is the charge of the particle and $\mathbf{k}$ is the curvature vector defined such that

$$\hat{\mathbf{b}}\times\mathbf{k} = \nabla\times\hat{\mathbf{b}} - (\hat{\mathbf{b}}\cdot\nabla\times\hat{\mathbf{b}})\hat{\mathbf{b}}. \tag{121}$$

We have also introduced $\Omega_\chi$,

$$\Omega_\chi = v_\parallel\hat{\mathbf{b}}\cdot\nabla\chi = \frac{v_\parallel c}{qR_0(1+\varepsilon\cos(\theta))}, \tag{122}$$

where for convenience we have defined the function

$$c(r) = \frac{1}{\sqrt{1+\varepsilon^2/\bar{q}^2}}. \tag{123}$$

Using these equations of motion, we find that

$$\begin{aligned}\dot{\phi} &= \frac{dq}{d\psi}\Omega_\chi|_{\psi_0}\psi_1 + \mathbf{v}_D\cdot\nabla\phi|_{\psi_0} - q\mathbf{v}_D\cdot\nabla\chi|_{\psi_0} + q\frac{d\chi}{dt}\\ &= \omega_d + q\frac{d\chi}{dt},\end{aligned} \tag{124}$$

where $\omega_d$ is the term that corresponds to the deviation from the magnetic field line.

To proceed with the calculation, we define a time average using the field-line-following orbit as reference. For trapped particles, the time average of a function $F(\varepsilon_\parallel, \chi)$, also called the bounce average, is

$$\langle F\rangle = \frac{2\pi}{\omega_b}\int_{-\chi_b}^{\chi_b}\frac{d\chi}{|\Omega_\chi|}\left(F(\varepsilon_\parallel,\chi)+F(-\varepsilon_\parallel,\chi)\right). \tag{125}$$

For passing particles, we instead have the transit average

$$\langle F\rangle = \frac{2\pi}{\omega_t}\int_{-\pi}^{\pi}\frac{d\chi}{|\Omega_\chi|}F(\varepsilon_\parallel,\chi). \tag{126}$$

To simplify our calculations, we perform a change of variables from $\chi$ to $\theta$,

$$\int_{-\chi_b}^{\chi_b}\frac{d\chi}{|\Omega_\chi|} = \int_{-\theta_b}^{\theta_b} d\theta\frac{\bar{q}R_0}{c|v_\parallel|}. \tag{127}$$

We then find that for trapped particles

$$\langle\dot{\phi}\rangle = \langle\omega_d\rangle, \tag{128}$$

and for passing particles

$$\langle\dot{\phi}\rangle = \langle\omega_d\rangle + q\varepsilon_\parallel\omega_t, \tag{129}$$

where $\omega_t$ is the transit frequency. The additional term for passing particles accounts for the change in the toroidal coordinate after a $2\pi$ change in the poloidal angle. Thus, it is associated with the lowest-order transit motion examined in Secs. III and IV.

The explicit expression for $\omega_d$ is

$$\omega_d = \frac{m\bar{q}v_\parallel^2}{eB_0r^2}\left(c^2\hat{s}+\frac{1}{2}I_1\right)+\frac{E\bar{q}}{eB_0r^2}I_2. \tag{130}$$

We have defined the magnetic shears,

$$s = \frac{r}{\bar{q}}\frac{d\bar{q}}{dr}, \tag{131}$$

$$\hat{s} = \frac{r}{q}\frac{dq}{dr} = \left(s+\frac{\varepsilon^2}{1-\varepsilon^2}\right), \tag{132}$$

and the functions $I_1(\theta)$, $I_2(\theta)$

$$I_1 = 1-(s-1)(1-c^2)+\frac{1-2c^2}{1+\varepsilon\cos(\theta)} + \frac{c^2\varepsilon^2\sin^2(\theta)}{(1-\varepsilon^2)(1+\varepsilon\cos(\theta)^2)}, \tag{133}$$

$$I_2 = (s-1)(1-c^2)+1-\frac{1}{1+\varepsilon\cos(\theta)} + \frac{c^2\varepsilon^2\sin^2(\theta)}{(1-\varepsilon^2)(1+\varepsilon\cos(\theta)^2)}. \tag{134}$$

We can then perform a time average, noting that

$$v_\parallel^2 = \frac{4E\varepsilon}{m}\frac{\kappa^2-\sin^2(\theta/2)}{1+\varepsilon\cos(\theta)}. \tag{135}$$

Using the elliptic integrals tabulated in the Appendix, we obtain for trapped particles

$$\langle\omega_d\rangle = \frac{E\bar{q}}{eB_0r^2F_1}\Bigg[\left((s-1)(1-c^2)+1\right)F_1 - F_2 + \frac{c^2\varepsilon^2}{1-\varepsilon^2}F_3 + 2\varepsilon\left(2\hat{s}+1-(s-1)(1-c^2)\right)F_4 + 2\varepsilon(1-2c^2)F_5+\frac{2c^2\varepsilon^3}{1-\varepsilon^2}F_6\Bigg]. \tag{136}$$

For passing particles, we instead have

$$\langle\omega_d\rangle = \frac{E\bar{q}}{eB_0r^2G_1}\Bigg[\left((s-1)(1-c^2)+1\right)G_1 - G_2 + \frac{c^2\varepsilon^2}{1-\varepsilon^2}G_3 + 2\varepsilon\left(2\hat{s}+1-(s-1)(1-c^2)\right)G_4 + 2\varepsilon(1-2c^2)G_5+\frac{2c^2\varepsilon^3}{1-\varepsilon^2}G_6\Bigg]. \tag{137}$$

We now take the limit that $\varepsilon$ is small. In this limit,

$$s \approx \hat{s}, \tag{138}$$

$$c \approx 1, \tag{139}$$

and the elliptic integrals simplify as outlined in the Appendix. We also see that the characteristic drift frequency is

$$\omega_{d,0} = \frac{E\bar{q}}{eB_0rR_0}\approx\frac{\omega_0^2\bar{q}^3}{\Omega_c\varepsilon}. \tag{140}$$

For trapped particles, we then obtain

$$\langle\omega_d\rangle \approx \frac{\omega_{d,0}\left((2+4\hat{s})E(\kappa)-(1+4\hat{s}(1-\kappa^2))K(\kappa)\right)}{K(\kappa)}, \tag{141}$$

which is the standard result. For passing particles, the drift frequency is approximately

$$\langle\omega_d\rangle \approx \frac{\omega_{d,0}\left(2\kappa^2(1+2\hat{s})E(\kappa^{-1})-(2\kappa^2-1)K(\kappa^{-1})\right)}{K(\kappa^{-1})}. \tag{142}$$

Figures 16–18 compare the toroidal drift frequency for various values of $\varepsilon$ and $\hat{s}$. We note that we only plot up to each individual case's respective upper bound for $\kappa$ shown in Eq. (22). We see that the exact formula is well approximated by the small $\varepsilon$ limit for $\varepsilon \lesssim 0.3$. However, finite inverse aspect ratio effects become important for larger values of $\varepsilon$. Departure from the small $\varepsilon$ limit is more noticeable for low magnetic shear in Fig. 17 and negative magnetic shear in Fig. 18. One can show that the magnetic drift at the trapped-passing boundary is

$$\langle\omega_d\rangle(\kappa=1) = \omega_{d,0}\left(\frac{(s-1)(1-c^2)}{\varepsilon}-\frac{1}{1-\varepsilon}\right). \tag{143}$$

The $\varepsilon$ dependence results in a large discrepancy for large values of $\varepsilon$. For the negative magnetic shear case, there is thus a large spike near the trapped-passing boundary for large values of $\varepsilon$ that is absent for small values of $\varepsilon$.

Figures 16–18 demonstrate that not only does a sign reversal for the toroidal drift frequency take place, but the value of $\kappa$ corresponding to the drift reversal is dependent on both $\varepsilon$ and $\hat{s}$. Figure 19 plots

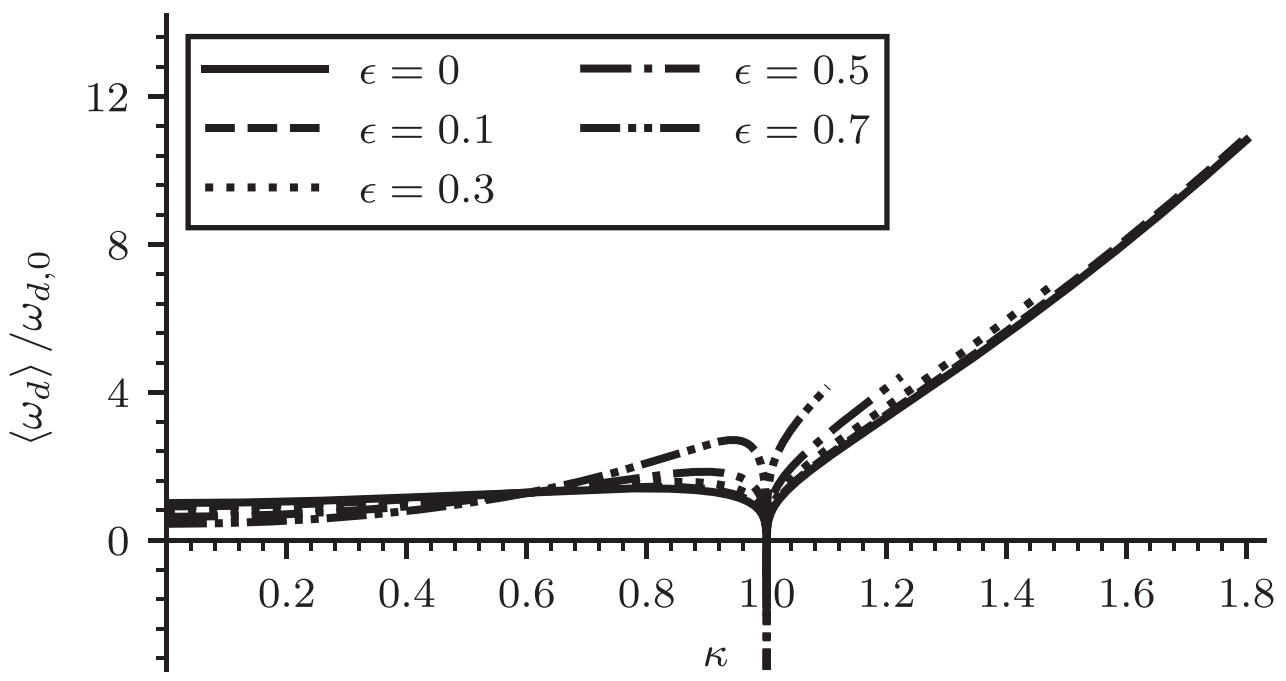

**FIG. 16.** Normalized toroidal drift frequency $\langle \omega_d \rangle / \omega_{d,0}$ vs the trapping parameter $\kappa$ for $\hat{s} = 1,\ \bar{q} = 2$. The trapped-passing boundary can easily be seen at $\kappa = 1$, around which the toroidal drift frequency changes sign. Discrepancies from the small $\varepsilon$ limit are large only for $\varepsilon = 0.7$.

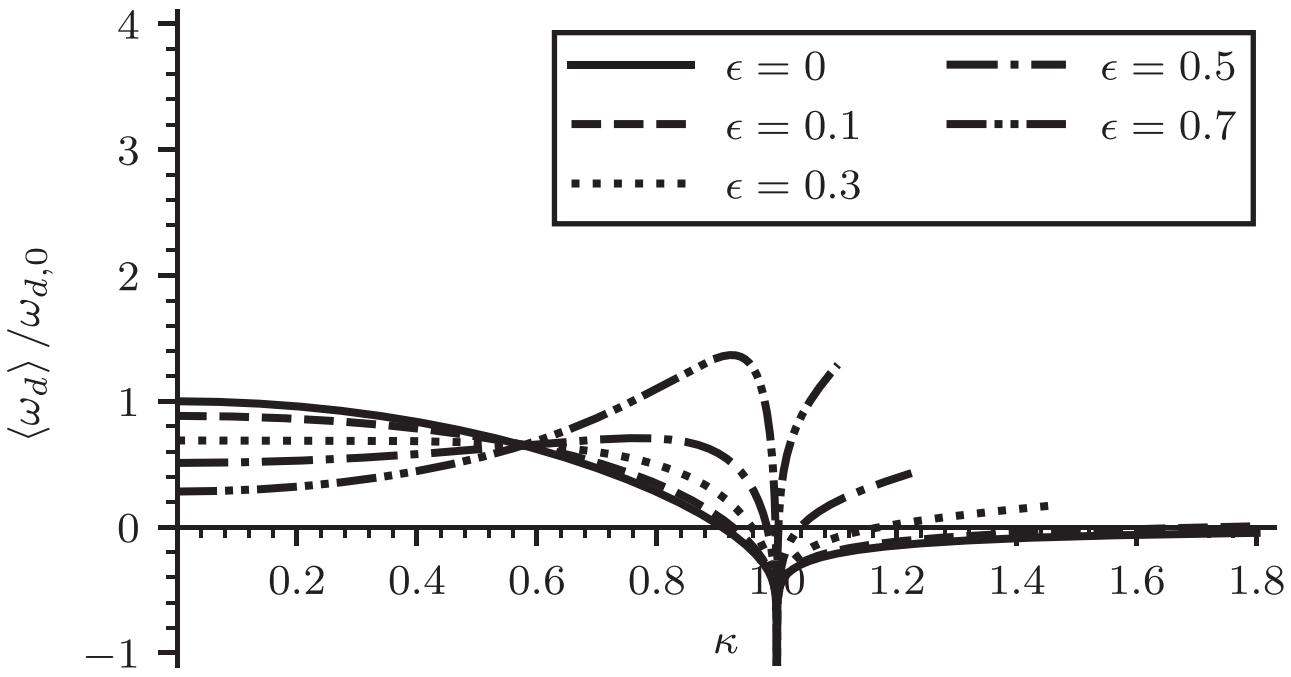

**FIG. 17.** Normalized toroidal drift frequency $\langle \omega_d \rangle / \omega_{d,0}$ vs the trapping parameter $\kappa$ for $\hat{s} = 0,\ \bar{q} = 2$. The trapped-passing boundary can easily be seen at $\kappa = 1$, around which the toroidal drift frequency changes sign. Discrepancies from the small $\varepsilon$ limit become large for $\varepsilon = 0.5,\ 0.7$.

the value of $\kappa$ at which the reversal takes place as a function of $\hat{s}$. The $\varepsilon = 0.5$ case diverges from the small $\varepsilon$ limit, and the extreme $\varepsilon = 0.7$ case shows markedly different behavior. The difference is not noticeable for large, positive $\hat{s}$.

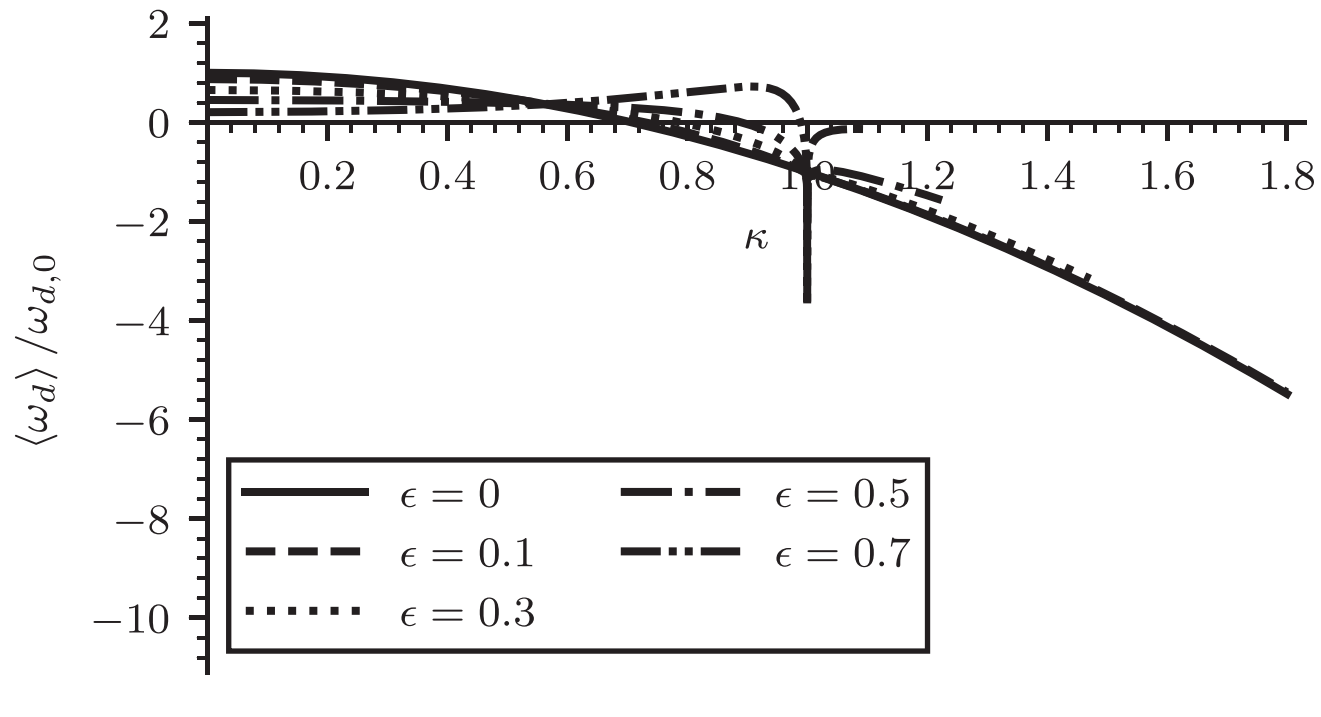

**FIG. 18.** Normalized toroidal drift frequency $\langle \omega_d \rangle / \omega_{d,0}$ vs the trapping parameter $\kappa$ for $\hat{s} = -0.5,\ \bar{q} = 2$. For small $\varepsilon$, the toroidal drift frequency undergoes sign reversal relatively far from the trapped-passing boundary. The behavior near the trapped-passing boundary at $\kappa = 1$ noticeably differs from the small $\varepsilon$ limit for $\varepsilon = 0.7$.

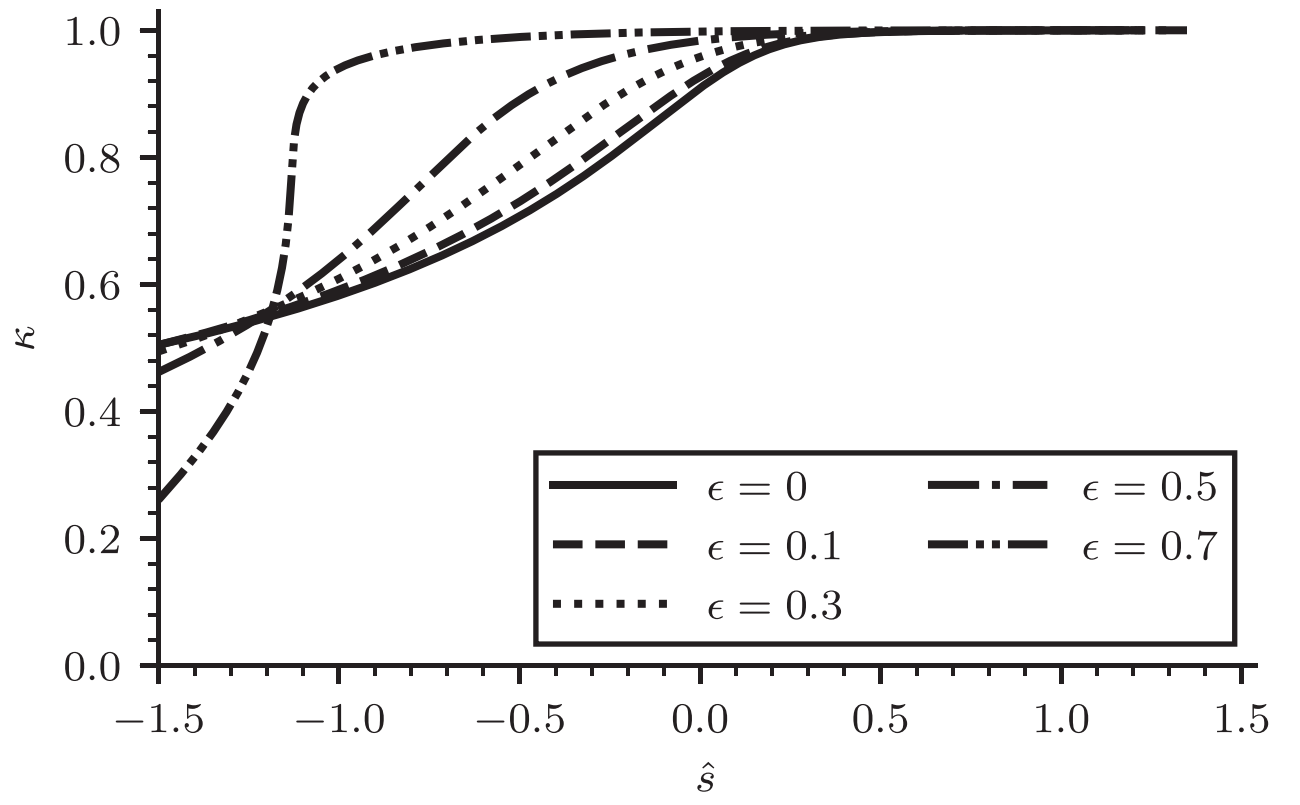

**FIG. 19.** The trapping parameter $\kappa$ at which the toroidal drift frequency changes sign vs the magnetic shear $\hat{s}$ for $\bar{q} = 2$. A decrease in magnetic shear corresponds to drift reversal occurring at a lower value of $\kappa$. Discrepancies from the small $\varepsilon$ limit are large for $\varepsilon = 0.5,\ 0.7$.

## VII. CONCLUSIONS

The calculations in this paper demonstrate that analytical formulas for the bounce-transit motion in a concentric circular geometry can be obtained without a small $\varepsilon$ approximation. In addition, the standard small $\varepsilon$ formulas can be easily obtained from the exact formulas; this method guarantees that a consistent ordering is applied in deriving the approximate expressions. A comparison using the analytical expressions reveals that the approximation is well suited for $\varepsilon \lesssim 0.3$. The approximate expressions for the longitudinal invariant and bounce-transit frequency are also robust for larger values of $\varepsilon$. However, the toroidal drift frequency is not well approximated for large $\varepsilon$, even absent the inclusion of shaping effects. Of course, one must keep in mind that the concentric circular geometry is itself an approximation which can significantly break down in the edge region. However, additional shaping effects such as the Shafranov shift could also be perturbatively included while retaining a finite $\varepsilon$.

Our analysis demonstrates that in cases where a concentric circular geometry is justified, applying a small $\varepsilon$ approximation to the geometry itself and subsequent calculations is also often justified. The formulas derived in this paper can be utilized while restricting oneself to a concentric circular geometry if additional accuracy is required, which may prove useful for spherical tokamaks.

## ACKNOWLEDGMENTS

We gratefully thank Jonathan Citrin and Clarisse Bourdelle for valuable discussions.

## DATA AVAILABILITY

Data sharing is not applicable to this article as no new data were created or analyzed in this study.

## APPENDIX A: ELLIPTIC INTEGRAL NOTATION

We clarify the notation used for elliptic integrals, as different sources will often have conflicting notation. The complete elliptic integral of the first kind is defined as

$$K(k) = \int_0^{\pi/2} \frac{d\varphi}{\sqrt{1 - k^2 \sin^2(\varphi)}}, \tag{A1}$$

where $k$ is the modulus. The complete elliptic integral of the second kind is

$$E(k) = \int_0^{\pi/2} d\varphi \sqrt{1 - k^2 \sin^2(\varphi)}. \tag{A2}$$

The incomplete integral of the third kind is

$$\Pi(\varphi, n, k) = \int_0^{\varphi} \frac{d\varphi'}{\sqrt{1 - k^2 \sin^2(\varphi')}\left(1 - n \sin^2(\varphi')\right)}, \tag{A3}$$

where $n$ is the characteristic. The integral is called complete when $\varphi = \pi/2$,

$$\Pi(n, k) = \int_0^{\pi/2} \frac{d\varphi}{\sqrt{1 - k^2 \sin^2(\varphi)}\left(1 - n \sin^2(\varphi)\right)}. \tag{A4}$$

We note that $n$ is allowed to be positive or negative.

For trapped particles, the modulus $k$ and characteristic $n$ are

$$k_b = \kappa \sqrt{\frac{1 - \varepsilon}{1 + \varepsilon - 2\varepsilon\kappa^2}}, \tag{A5}$$

$$n_b = \frac{-2\varepsilon\kappa^2}{1 + \varepsilon - 2\varepsilon\kappa^2}. \tag{A6}$$

For passing particles, the modulus and characteristic are

$$k_t = \frac{1}{\kappa} \sqrt{\frac{1 + \varepsilon - 2\varepsilon\kappa^2}{1 - \varepsilon}}, \tag{A7}$$

$$n_t = \frac{-2\varepsilon}{1 - \varepsilon}. \tag{A8}$$

## APPENDIX B: SPECIFIC ELLIPTIC INTEGRALS

The calculation of several elliptic integrals is necessary to carry out the analysis in Sec. VI. These integrals are tabulated below with the help of Refs. 22–24. We first list those corresponding to trapped particles

$$F_1(\varepsilon, \kappa) = \int_0^{\theta_b} d\theta \sqrt{\frac{1 + \varepsilon \cos(\theta)}{\kappa^2 - \sin^2(\theta/2)}} = \frac{2(1 + \varepsilon)\Pi(n_b, k_b)}{\sqrt{1 + \varepsilon - 2\varepsilon\kappa^2}}, \tag{B1}$$

$$F_2(\varepsilon, \kappa) = \int_0^{\theta_b} d\theta \sqrt{\frac{1 + \varepsilon \cos(\theta)}{\kappa^2 - \sin^2(\theta/2)}} \frac{1}{1 + \varepsilon \cos(\theta)} = \frac{2K(k_b)}{\sqrt{1 + \varepsilon - 2\varepsilon\kappa^2}}, \tag{B2}$$

$$F_3(\varepsilon, \kappa) = \int_0^{\theta_b} d\theta \sqrt{\frac{1 + \varepsilon \cos(\theta)}{\kappa^2 - \sin^2(\theta/2)}} \frac{\sin^2(\theta)}{(1 + \varepsilon \cos(\theta))^2} = \frac{2\left((1 - \varepsilon)K(k_b) + 2\varepsilon E(k_b) - (1 + \varepsilon)\Pi(n_b, k_b)\right)}{\varepsilon^2 \sqrt{1 + \varepsilon - 2\varepsilon\kappa^2}}, \tag{B3}$$

$$F_4(\varepsilon, \kappa) = \int_0^{\theta_b} d\theta \sqrt{\frac{\kappa^2 - \sin^2(\theta/2)}{1 + \varepsilon \cos(\theta)}} = \frac{\sqrt{1 + \varepsilon - 2\varepsilon\kappa^2}}{\varepsilon} \left(\frac{(1 + \varepsilon)\Pi(n_b, k_b)}{1 + \varepsilon - 2\varepsilon\kappa^2} - K(k_b)\right), \tag{B4}$$

$$F_5(\varepsilon, \kappa) = \int_0^{\theta_b} d\theta \sqrt{\frac{\kappa^2 - \sin^2(\theta/2)}{1 + \varepsilon \cos(\theta)}} \frac{1}{1 + \varepsilon \cos(\theta)} = \frac{2\sqrt{1 + \varepsilon - 2\varepsilon\kappa^2}}{1 - \varepsilon^2} \left(E(k_b) - (1 - k_b^2)K(k_b)\right), \tag{B5}$$

$$F_6(\varepsilon, \kappa) = \int_0^{\theta_b} d\theta \sqrt{\frac{\kappa^2 - \sin^2(\theta/2)}{1 + \varepsilon \cos(\theta)}} \frac{\sin^2(\theta)}{(1 + \varepsilon \cos(\theta))^2} = \frac{(1 + \varepsilon)\left(a_b K(k_b) + b_b \Pi(n_b, k_b)\right) + c_b E(k_b)}{3\varepsilon^3 (1 - \varepsilon^2)\sqrt{1 + \varepsilon - 2\varepsilon\kappa^2}}. \tag{B6}$$

We have defined additional coefficients for $F_6$,

$$a_b = 3 - 6\varepsilon + (4\kappa^2 - 1)\varepsilon^2, \tag{B7}$$

$$b_b = -3(1 - \varepsilon^2), \tag{B8}$$

$$c_b = 2\varepsilon(3 - (4\kappa^2 - 2) - \varepsilon^2). \tag{B9}$$

For passing particles, we have instead

$$G_1(\varepsilon, \kappa) = \int_0^{\pi} d\theta \sqrt{\frac{1 + \varepsilon \cos(\theta)}{\kappa^2 - \sin^2(\theta/2)}} = \frac{2(1 + \varepsilon)\Pi(n_t, k_t)}{\kappa\sqrt{1 - \varepsilon}}, \tag{B10}$$

$$G_2(\varepsilon, \kappa) = \int_0^{\pi} d\theta \sqrt{\frac{1 + \varepsilon \cos(\theta)}{\kappa^2 - \sin^2(\theta/2)}} \frac{1}{1 + \varepsilon \cos(\theta)} = \frac{2K(k_t)}{\kappa\sqrt{1 - \varepsilon}}, \tag{B11}$$

$$G_3(\varepsilon, \kappa) = \int_0^{\pi} d\theta \sqrt{\frac{1 + \varepsilon \cos(\theta)}{\kappa^2 - \sin^2(\theta/2)}} \frac{\sin^2(\theta)}{(1 + \varepsilon \cos(\theta))^2} = \left(k_t^2 \left(K(k_t) - \Pi(n_t, k_t)\right) + \frac{n_t (K(k_t) - E(k_t))}{1 - n_t}\right) \times \frac{2\kappa (1 - \varepsilon)^{3/2}(1 - n_t)}{\varepsilon^2 (1 + \varepsilon - 2\varepsilon\kappa^2)}, \tag{B12}$$

$$G_4(\varepsilon, \kappa) = \int_0^{\pi} d\theta \sqrt{\frac{\kappa^2 - \sin^2(\theta/2)}{1 + \varepsilon \cos(\theta)}} = \frac{(1 + \varepsilon)\Pi(n_t, k_t) - (1 + \varepsilon - 2\varepsilon\kappa^2)K(k_t)}{\varepsilon\kappa\sqrt{1 - \varepsilon}}, \tag{B13}$$

$$G_5(\varepsilon, \kappa) = \int_0^{\pi} d\theta \sqrt{\frac{\kappa^2 - \sin^2(\theta/2)}{1 + \varepsilon \cos(\theta)}} \frac{1}{1 + \varepsilon \cos(\theta)} = \frac{2\kappa E(k_t)}{\sqrt{1 - \varepsilon}(\varepsilon + 1)}, \tag{B14}$$

$$G_6(\varepsilon, \kappa) = \int_0^{\pi} d\theta \sqrt{\frac{\kappa^2 - \sin^2(\theta/2)}{1 + \varepsilon \cos(\theta)}} \frac{\sin^2(\theta)}{(1 + \varepsilon \cos(\theta))^2} = \frac{(1 + \varepsilon)\left(a_t K(k_t) + b_t \Pi(n_t, k_t)\right) + c_t E(k_t)}{3\varepsilon^3 \kappa\sqrt{1 - \varepsilon}(1 + \varepsilon)(1 + \varepsilon - 2\varepsilon\kappa^2)}. \tag{B15}$$

The coefficients for $G_6$ are

$$a_t = 3 - (12\kappa^2 - 6)\varepsilon + (8\kappa^4 - 8\kappa^2 + 3)\varepsilon^2, \tag{B16}$$

$$b_t = -3(1 - 2(\kappa^2 - 1)\varepsilon - (2\kappa^2 - 1)\varepsilon^2), \tag{B17}$$

$$c_t = 2\varepsilon\kappa^2(3 - (4\kappa^2 - 2)\varepsilon - \varepsilon^2). \tag{B18}$$

We also tabulate the small $\varepsilon$ expansions used in Sec. VI,

$$F_1 \approx 2K(\kappa) + \varepsilon(2E(\kappa) - K(\kappa)), \quad \text{(B19)}$$

$$F_2 \approx 2K(\kappa) - \varepsilon(2E(\kappa) - K(\kappa)), \quad \text{(B20)}$$

$$F_3 \approx \frac{8}{3}\left((2\kappa^2 - 1)E(\kappa) - (1 - \kappa^2)K(\kappa)\right), \quad \text{(B21)}$$

$$F_4 \approx 2\left(E(\kappa) - (1 - \kappa^2)K(\kappa)\right), \quad \text{(B22)}$$

$$F_5 \approx 2\left(E(\kappa) - (1 - \kappa^2)K(\kappa)\right), \quad \text{(B23)}$$

$$F_6 \approx -\frac{8}{15}\left((2 - 3\kappa^2 + \kappa^4)K(\kappa) - 2(1 - \kappa^2 + \kappa^4)E(\kappa)\right), \quad \text{(B24)}$$

$$G_1 \approx \frac{2}{\kappa}K(\kappa^{-1}) + \frac{\varepsilon}{\kappa}\left(2E(\kappa^{-1}) - (2\kappa^2 - 1)K(\kappa^{-1})\right), \quad \text{(B25)}$$

$$G_2 \approx \frac{2}{\kappa}K(\kappa^{-1}) - \frac{\varepsilon}{\kappa}\left(2E(\kappa^{-1}) - (2\kappa^2 - 1)K(\kappa^{-1})\right), \quad \text{(B26)}$$

$$G_3 \approx \frac{8}{3}\kappa\left((2\kappa^2 - 1)E(\kappa^{-1}) - 2(\kappa^2 - 1)K(\kappa^{-1})\right), \quad \text{(B27)}$$

$$G_4 \approx 2\kappa E(\kappa^{-1}), \quad \text{(B28)}$$

$$G_5 \approx 2\kappa E(\kappa^{-1}), \quad \text{(B29)}$$

$$G_6 \approx \frac{8}{15}\kappa\left(2(\kappa^4 - \kappa^2 + 1)E(\kappa^{-1}) - (\kappa^2 - 1)(2\kappa^2 - 1)K(\kappa^{-1})\right). \quad \text{(B30)}$$